  \documentclass{nature_armin2}
\usepackage{longtable}
\usepackage{lscape}

\newcommand{\gps}{\ensuremath{g_{\rm P1}}}
\newcommand{\rps}{\ensuremath{r_{\rm P1}}}
\newcommand{\ips}{\ensuremath{i_{\rm P1}}}
\newcommand{\zps}{\ensuremath{z_{\rm P1}}}
\newcommand{\yps}{\ensuremath{y_{\rm P1}}}

\newcommand{\PS}{\protect \hbox {Pan-STARRS1}}

\newcommand{\etal}{et al.~}
\newcommand{\nuv}{\ensuremath{NUV}}
\newcommand{\aplt} {\ {\raise-.5ex\hbox{$\buildrel<\over\sim$}}\ } 
\newcommand{\apgt} {\ {\raise-.5ex\hbox{$\buildrel>\over\sim$}}\ }

\newcommand\ion[2]{#1$\;${%
\ifx\@currsize\normalsize\small \else
\ifx\@currsize\small\footnotesize \else
\ifx\@currsize\footnotesize\scriptsize \else
\ifx\@currsize\scriptsize\tiny \else
\ifx\@currsize\large\normalsize \else
\ifx\@currsize\Large\large
\fi\fi\fi\fi\fi\fi
\rmfamily#2}\relax}%

\usepackage{epstopdf}
\usepackage{amsmath}
\usepackage{rotating}

\usepackage{color}

\def\reff@jnl#1{{\rm#1\/}}
\def\apj{\reff@jnl{Astrophys. J.}}
\def\apjl{\reff@jnl{Astrophys. J.}}               
\def\aj{\reff@jnl{Astron. J.}}                  
\def\araa{\reff@jnl{Annu. Rev. Astron. Astr.}}            
\def\apjs{\reff@jnl{Astrophys. J. Sup.}}              
\def\aap{\reff@jnl{Astron. Astrophys.}}               
\def\mnras{\reff@jnl{Mon. Not. R. Astron. Soc.}}      
\def\prd{\reff@jnl{Phys. Rev. D}}         
\def\prl{\reff@jnl{Phys. Rev. Lett.}}      
\def\pasp{\reff@jnl{Publ. Astron. Soc. Pac.}}              
\def\nat{\reff@jnl{Nature}}             

\def\arcmin{\hbox{$^\prime$}}
\def\arcsec{\hbox{$^{\prime\prime}$}}

\def\snid{\ifmmode{\rm \tt SNID}\else{\tt SNID}\fi}
\def\dm15{\ifmmode{\Delta m_{15}}\else{$\Delta m_{15}$}\fi}

\def\magarcsec2{\ \rm{mag \ arcsec}^{-2}}
\def\ly{\ifmmode{\rm{ly}}\else{ly\fi}}

\title{An ultraviolet-optical flare from the tidal disruption of a helium-rich stellar core}

\author{
S. Gezari$^{1}$,
R. Chornock$^{2}$,
A. Rest$^{3}$,
M. E. Huber$^{4}$,
K. Forster$^{5}$,
E. Berger$^{2}$,
P. J. Challis$^{2}$,
J. D. Neill$^{5}$,
D. C. Martin$^{5}$,
T. Heckman$^{1}$,
A. Lawrence$^{6}$,
C. Norman$^{1}$,
G. Narayan$^{2}$,
R. J. Foley$^{2}$,
G. H. Marion$^{2}$,
D. Scolnic$^{1}$,
L. Chomiuk$^{2}$,
A. Soderberg$^{2}$,
K. Smith$^{7}$,
R. P. Kirshner$^{2}$,
A. G. Riess$^{1}$,
S. J. Smartt$^{7}$,
C.W. Stubbs$^{2}$,
J.L. Tonry$^{4}$,
W. M. Wood-Vasey$^{8}$,
W. S. Burgett$^{4}$,
K. C. Chambers$^{4}$, 
T. Grav$^{9}$,
J. N. Heasley$^{4}$,
N. Kaiser$^{4}$,
R.-P. Kudritzki$^{4}$,
E. A. Magnier$^{4}$,
J. S. Morgan$^{4}$, \&
P. A. Price$^{10}$
}

\begin{document}

\maketitle

\begin{affiliations}
\item Department of Physics and Astronomy, Johns Hopkins University, 3400 North Charles Street, Baltimore, MD 21218, USA.
\item Harvard-Smithsonian Center for Astrophysics, 60 Garden Street, Cambridge, MA 02138, USA.
\item Space Telescope Science Institute, 3700 San Martin Drive, Baltimore, MD 21218, USA.
\item Institute for Astronomy, University of Hawaii, 2680 Woodlawn Drive, Honolulu HI 96822, USA.
\item California Institute of Technology, 1200 East California Blvd., Pasadena, CA 91125, USA.
\item Institute for Astronomy, University of Edinburgh Scottish Universities Physics Alliance, Royal Observatory, Blackford Hill, Edinburgh, EH9 3HJ, UK.
\item Astrophysics Research Centre, School of Mathematics and Physics, Queen's University Belfast, Belfast, BT7 1NN, UK.
\item Pittsburgh Particle Physics, Astrophysics, and Cosmology Center, Department of Physics and Astronomy, University of Pittsburgh, 3941 O'Hara Street, Pittsburgh, PA 15260, USA.
\item Planetary Science Institute, 1700 East Fort Lowell, Tucson, AZ 85719, USA
\item Department of Astrophysical Sciences, Princeton University, Princeton, NJ 08544, USA.
\end{affiliations}


\begin{abstract} 
The flare of radiation from the tidal
disruption and accretion of a star can be used 
as a marker for supermassive black holes that
otherwise lie dormant and undetected 
in the centres of distant galaxies\cite{Rees1988}.
Previous candidate flares\cite{Komossa1999, Komossa2004, Esquej2008,
Gezari2009, vanVelzen2011} have had declining light curves in good
agreement with expectations, but with poor constraints on the time
of disruption and the type of star disrupted, because the rising emission
was not observed.  Recently, two
`relativistic' candidate tidal disruption events were discovered, each of 
whose extreme X-ray
luminosity and synchrotron
radio emission were interpreted as the onset of emission from a relativistic jet
\cite{Bloom2011, Burrows2011,
Zauderer2011, Cenko2011}.   
Here we report the discovery of a luminous
ultraviolet-optical flare from the nuclear region of an inactive
galaxy at a redshift of $0.1696$.  The observed continuum
is cooler than expected for a simple accreting debris disk, but the 
well-sampled rise and decline of its light curve follows 
the predicted mass accretion rate, 
and can be modelled to determine
the time of disruption to an accuracy of two days.  The black hole 
has a mass of about 2 million solar masses, 
modulo a factor dependent on the mass and radius of the star disrupted.  
On the basis of the spectroscopic signature of ionized helium from the unbound
debris, we determine that the disrupted star was a helium-rich stellar core.  
\end{abstract}

When the pericenter of a star's orbit ($R_{\rm p}$) passes within the tidal disruption radius of a massive black hole, $R_{\rm T} \approx R_{\star}(M_{\rm BH}/M_{\star})^{1/3}$, tidal forces overcome the binding energy of the star, which breaks up with roughly half of the stellar debris remaining bound to the black hole and the rest being ejected at high velocity\cite{Rees1988}.  For black holes above a critical mass, $M_{\rm crit} \approx 10^{8} r_{\star}^{3/2}m_{\star}^{-1/2} M_{\odot}$ (where $r_{\star} = R_{\star}/R_{\odot}$ and $m_{\star} = M_{\star}/M_{\odot}$), the star becomes trapped within the event horizon of the black hole before being disrupted.
The mass accretion rate ($\dot{M}$) in a tidal disruption event (TDE) can be calculated directly from the orbital return-times of the bound debris\cite{Rees1988,Phinney1989, Evans1989}.  
For the simplest case of a star of uniform density this yields, $\dot{M} = \frac{2}{3} (\frac{fM_{\star}}{t_{\rm min}}) (\frac{t}{t_{\rm min}})^{-5/3}$, where $f$ is the fraction of the star accreted and $t_{\rm min}$ is the orbital period of the most tightly bound debris and, therefore, the time delay between the time of disruption and the start of the flare, which scales as $M_{\rm BH}^{1/2}M_{\star}^{-1}R_{\star}^{3/2}$ for $R_{\rm p} = R_{\rm T}$.
The radiative output of the accreted debris is less certain, and depends on the ratio of the accretion rate to the Eddington rate\cite{Ulmer1999}.

The optical transient, PS1-10jh ($\alpha_{\rm J2000} = 16 $h$ 09 $min$ 28.296 $s$, \delta_{\rm J2000} = +53\deg 40\arcmin 23.52\arcsec$), was discovered on 2010 May 31.45 UT in the \PS\cite{Kaiser2010} (PS1) Medium Deep Survey by our two independent image-differencing pipelines.  The densely sampled (cadence,$ \Delta \sim 3$ d) optical light curves of PS1-10jh in the \gps, \rps, \ips, and \zps\ bands (Supplementary Information) follow the rise of the transient to its peak in the \gps\ band on 2010 July 12.31 UT and its subsequent decline until 2011 September 1.24 UT (Supplementary Table 1).  PS1-10jh was discovered independently as a transient, near-ultraviolet (\nuv) source at the 20$\sigma$ level by the \textsl{GALEX}\cite{Martin2005} Time Domain Survey (TDS) on 2010 June 17.68 UT within $2.5 \pm 3.0$ arcsec of the PS1 location, and was detected in ten more epochs of TDS observations between then and 2011 June 10.68 UT (Supplementary Table 2).  No source is detected in a deep coadd of all the TDS epochs in 2009, with a 3$\sigma$ upper limit of $> 25.6$ mag implying a peak amplitude of variability in the $NUV$ of $> 6.4$ mag.  See the Supplementary Information for details on the PS1 and \textsl{GALEX} photometry.

\begin{figure}
\caption{
\begin{centering}
 \includegraphics[scale=0.75]{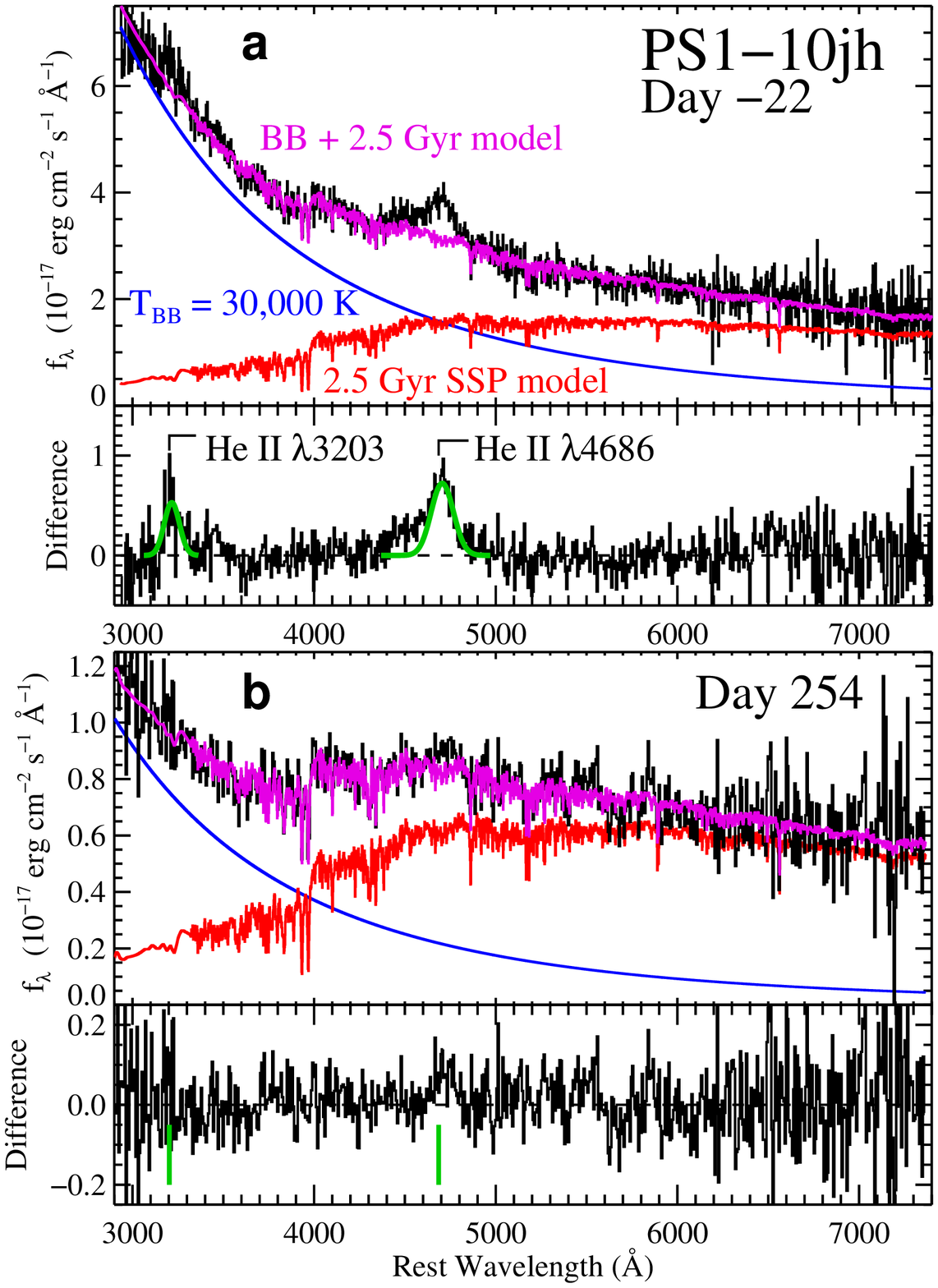}
\end{centering}
\\
Optical spectrum.  MMT optical spectra (black) of PS1-10jh obtained $-22$ ({\bf a}) and $+254$ ({\bf b}) rest-frame days from the peak, expressed in terms of flux density.  Each continuum is fitted with a combination (magenta) of a stellar population $2.5$ Gyr old and a fading blackbody with a temperature of $\sim 3 \times 10^{4}$ K determined from the ultraviolet-optical spectral energy distribution (SED).  \ion{He}{II} emission at $\lambda=4,686$\AA (Fowler series $n = 4 \rightarrow 3$) is detected above the continuum model and fitted with a Gaussian with a full-width at half-maximum of $9,000 \pm 700$ km s$^{-1}$ and $L = (9 \pm 1)\times10^{40}$ ergs s$^{-1}$ (plotted with a green line in the early epoch ({\bf a})).  Residual emission above the continuum model is also detected at $\sim 3,200$ \AA\, which is coincident with the location of the \ion{He}{II}$\lambda 3203$ (Fowler series $n = 5 \rightarrow 3$) line, and confirms the identification of \ion{He}{II} $\lambda=4,686$ \AA emission.  The observed flux ratio of \ion{He}{II} $\lambda=3,203$ emission to \ion{He}{II} $\lambda=4,686$\AA emission is $0.50 \pm 0.10$, measured using a Gaussian fit to the $\lambda = 3,203$\AA line with a width fixed to that of the $\lambda=4,686$\AA line, limits the internal extinction to $E(B-V) < 0.08$ mag (Supplementary Information). The \ion{He}{II} $\lambda 4,686$ line is still evident as an excess above the model in the later epoch ({\bf b}), but it has faded by a factor of $\sim 10$ since 22 rest-frame days before the peak, the same factor by which the ultraviolet continuum has faded during this time.  The absolute flux scaling in the later epoch is uncertain owing to obscuration by clouds on the date of the observation.  
\label{fig:spectra_early}}
\end{figure}

PS1-10jh is coincident with the centre of a galaxy within the 3$\sigma$ positional uncertainty (0.036 arcsec; Supplementary Information) with rest-frame $u$, $g$, $r$, $i$, and $z$ photometry from SDSS\cite{Aihara2011} and $K$ photometry from UKIDSS\cite{Lawrence2007} fitted with a galaxy template\cite{Blanton2007} with $M_{\rm stars} = (3.6 \pm 0.2) \times 10^{9} M_{\odot}$ and $M_{\rm r} = -18.7$ mag, where $M_{\rm stars}$ is the galaxy stellar mass and $M_{\rm r}$ is the absolute $r$-band magnitude.  The mass of the central black hole as determined indirectly from locally established scaling relations\cite{Haring2004} is $4^{+4}_{-2} \times 10^{6} M_{\odot}$.  
We obtained five epochs of optical spectroscopy at the location of PS1-10jh between 2010 June 16.33 and 2011 September 4.23 UT with the 6.5-m MMT (Supplementary Table 3).  The continua in the spectra 
are well modelled by the combination of a galaxy host at redshift $0.1696$ (luminosity distance, $d_{L} = 816$ Mpc) with a stellar population with an age of $1.4-5.0$ Gyr, depending on the chosen metallicity, and a fading hot blackbody component with $T_{\rm BB} \approx 3 \times 10^{4}$ K (Fig.~\ref{fig:spectra_early}).  

The spectra show no narrow emission lines that would be indicative of star formation or an active galactic nucleus (AGN).  
We obtained a 10 ks, 0.2-10 keV X-ray observation, using the \textsl{Chandra} X-ray Observatory, at the location of PS1-10jh on 2011 May 22.96 UT, and detected no source above the background with a 3$\sigma$ upper limit of $L_{X}(0.2-10$ keV)~$ < 5.8\times10^{41}$ ergs s$^{-1}$ for an unobscured AGN spectrum.  The X-ray faintness and extreme $NUV$ variability amplitude of PS1-10jh strongly disfavour its origin in an AGN, and its prolonged brightness in the ultraviolet strongly disfavours its origin in a supernova (Supplementary Information).

\begin{figure}
\caption{
 \includegraphics[scale=0.75]{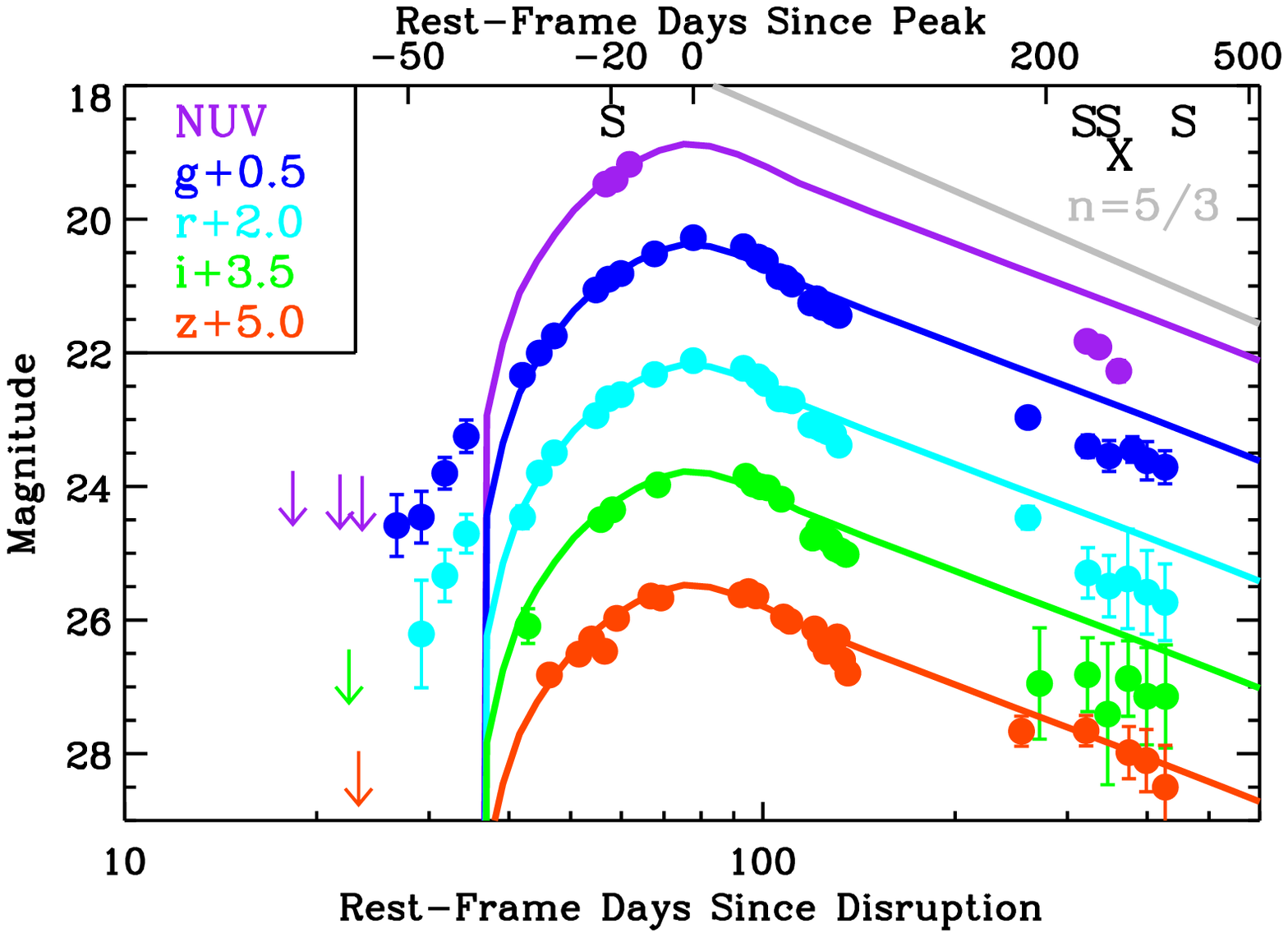}
\\
Ultraviolet-optical light curve.  The \textsl{GALEX} $NUV$ and PS1 \gps-, \rps-, \ips-, and \zps-band light curves of PS1-10jh (with the host galaxy flux removed), plotted against logarithmic time since the peak (top) and since the disruption (bottom).  The curves (shown with solid lines scaled to the flux in the \textsl{GALEX} and PS1 bands) were determined from the best fit of the \gps-band light curve to a numerical model\cite{Lodato2009} for the mass accretion rate of a tidally disrupted star with a polytropic exponent of $5/3$.  For each of the four optical bands, we independently performed a least-squares fit of the model for a $10^{6} M_{\odot}$ black hole to the light curve from $-36$ to $58$ rest-frame days from the peak, with the time of disruption, a vertical scaling factor, and a time stretch factor as free parameters.  The \textsl{GALEX} and PS1 photometry at $t > 240$ rest-frame days since the peak is shown binned in time in order to increase the signal-to-noise ratio.  The dates of multiple epochs of MMT spectroscopy are marked with an S, and the date of the \textsl{Chandra} X-ray observation is marked with an X.  The grey line shows an $n=5/3$ power-law decay from the peak.  Errors, 1$\sigma$; arrows, 3$\sigma$ upper limits.
\label{fig:rise}
}
\end{figure}

The rise and decay of the light curve of PS1-10jh is well described by numerical simulations for the mass return rate from a star that is tidally disrupted at $R_{\rm p} = R_{\rm T}$ with an internal structure parameterised by a polytropic exponent of $5/3$ characteristic of a fully convective star or a degenerate core\cite{Lodato2009} (Fig.~\ref{fig:rise}).  The decline from the peak is too steep to be fitted by simulations of a more centrally concentrated stellar structure, such as one that is characteristic of a solar-type star (Supplementary Information).  There are systematic differences between the light curve and the model during the early rise (more than $-44$ rest-frame days before the peak) and the late decay (more than $240$ rest-frame days after the peak) which could imply a stellar structure more complex than one described by a single polytrope.
The mass of the black hole is determined from the stretch factor of $1.38 \pm 0.03$ applied to fit the model of a $10^{6} M_{\odot}$ black hole to the light curve, which implies that the time of disruption was $76 \pm 2$ d before the peak and $M_{\rm BH} = (1.9 \pm 0.1) \times 10^{6} m_{\star}^{2}r_{\star}^{-3} M_{\odot}$.  

The most constraining property of PS1-10jh is the detection of very broad high-ionisation \ion{He}{II} emission at wavelengths of $\lambda = 4,686$ \AA (full-width at half-maximum, $9,000 \pm 700$ km s$^{-1}$) and $\lambda = 3,203$ \AA that fade in time along with the ultraviolet-optical continuum.  
The lack of Balmer line emission in the spectra requires an extremely low
hydrogen mass fraction, of $< 0.2$ (Supplementary Information), which cannot be found in the ambient interstellar medium or in a passive accretion disk.  This is the strongest evidence that PS1-10jh must be fuelled by the accretion of a star that has lost its hydrogen envelope, either through stellar winds or tidal interactions with the central supermassive black hole.  
The broad width of the line is also what is expected from the velocities of the most energetic unbound stellar debris in a tidal disruption event\cite{Strubbe2009}, that is $v_{\rm max} \sim 1\times10^{4} (M_{\rm BH}/10^{6} M_{\odot})^{1/6}(R_{\rm T}/R_{\rm p})r_{\star}^{-1/2}m_{\star}^{1/3}$ km s$^{-1}$.  

\begin{figure}
\caption{
\includegraphics[scale=0.75]{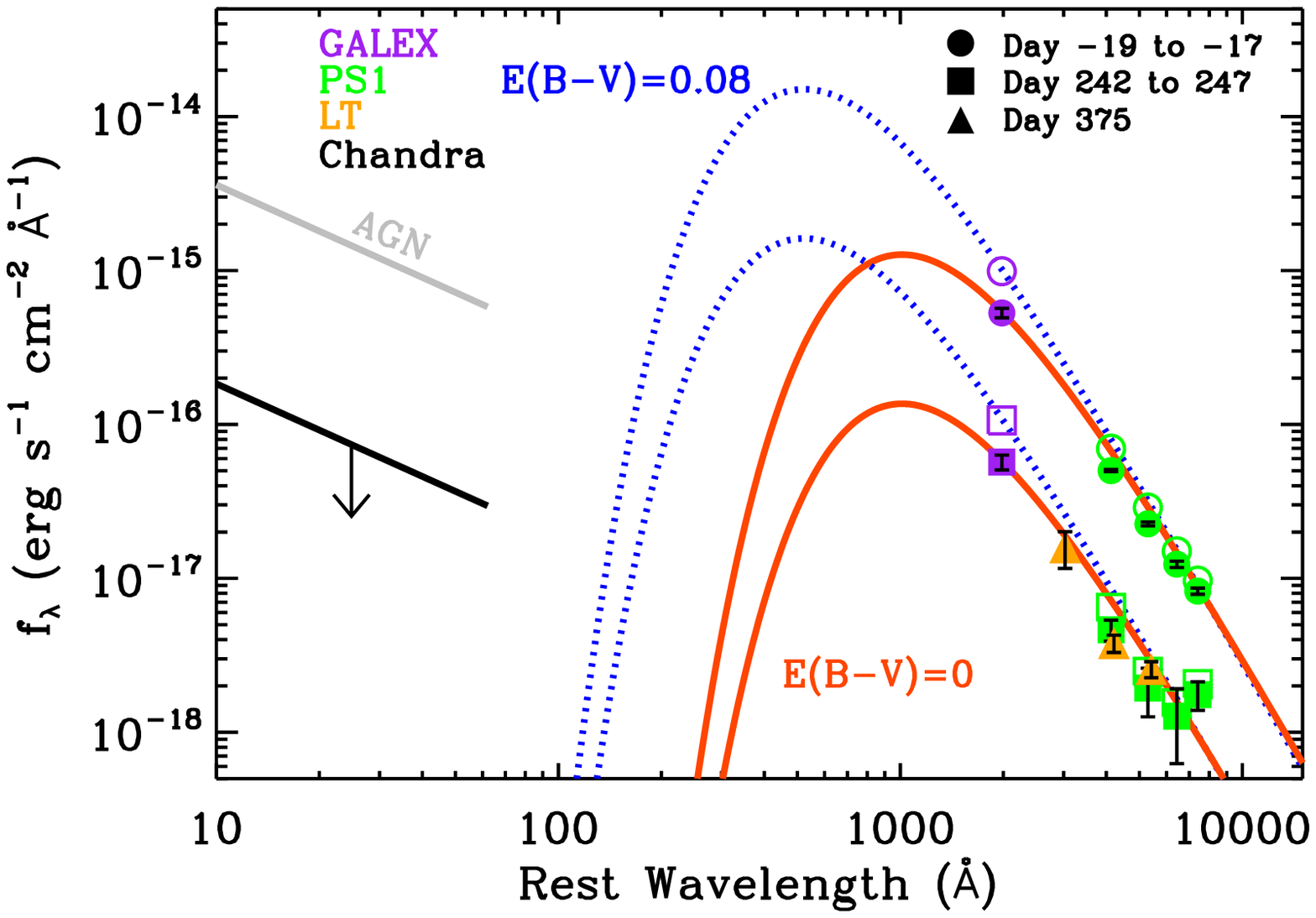}
\\
Spectral energy distribution.  SED of PS1-10jh during nearly simultaneous \textsl{GALEX} ultraviolet and PS1 optical observations (with the host galaxy flux removed) at two epochs (rest-frame days -19 to -17 and 242 to 247 from the peak of the flare).  Flux densities have been corrected for Galactic extinction of $E(B-V)=0.013$ mag.  The ultraviolet-optical SED from $-19$ to $247$ rest-frames days from the peak is fitted with a $2.9\times10^{4}$ K blackbody.  Orange solid lines show blackbodies with this temperature scaled to the $NUV$ flux densities for the respective epochs.   Open symbols show the \textsl{GALEX} and PS1 flux densities corrected for an internal extinction of $E(B-V) = 0.08$ mag, and dotted blue line shows the $5.5\times10^{4}$ K blackbody fit to the de-reddened flux densities for the respective epochs.  The upper limit from the Chandra observation on 2011 May 22.96 UT assuming a spectrum with a photon index of $\Gamma = 2$, typical of an AGN, is plotted with a thick black line.  The X-ray flux density expected from an unobscured AGN with a comparable $NUV$ flux is plotted for comparison with a thick grey line\cite{Steffen2006}.  Also shown are the $u$, $g$, and $r$ band flux densities measured from aperture photometry with the Liverpool Telescope\cite{Steele2004} on 2011 Sep 24 UT, after subtracting the host galaxy flux measured by SDSS.  Errors, 1$\sigma$.
\label{fig:sed}
}
\end{figure}

We measure the SED of the flare over time from the nearly simultaneous PS1 optical and \textsl{GALEX} ultraviolet observations (with the host galaxy flux removed; Fig.~\ref{fig:sed}).  The pre-peak SED is fitted with a blackbody with $T_{\rm BB} = (2.9 \pm 0.2)\times10^{4}$ K, consistent with the blackbody component seen in the spectra.  However, the temperature fit is very sensitive to internal extinction.  If we correct for the maximum internal extinction of $E(B-V) = 0.08$ mag allowed by the observed \ion{He}{II}$\lambda = 3,203$\AA, $\lambda = 4,686$\AA emission, the best-fit temperature increases to $(5.5 \pm 0.4) \times 10^{4}$ K.  In fact, we know that the photo-ionizing continuum must have $T_{\rm BB} \apgt 5 \times 10^{4}$ K $22$ rest-frame days before the peak in order to produce enough $\lambda < 228$ \AA\ photons to photoionise the \ion{He}{II} $\lambda=$4,686\AA line observed with a luminosity of ($9 \pm 1) \times 10^{40}$ ergs s$^{-1}$. 
The late-time SED can be fitted with the same temperature as the pre-peak SED. 
We note that the observed continuum temperature, and even the maximum temperature allowed by possible de-reddening, are considerably cooler than the temperature of $\approx 2.5 \times 10^{5} (M_{\rm BH}/10^{6} M_{\odot})^{1/12}r_{\star}^{-1/2}m_{\star}^{-1/6}$ K expected from material radiating at the Eddington limit at the tidal radius\cite{Ulmer1999}.  This discrepency is also seen in AGN\cite{Lawrence2011} and may imply that the continuum we see is due to reprocessing of some kind\cite{Loeb1997, Lawrence2011}

On the basis of the arguments above, we assume that the observed temperature is a lower limit, $T_{\rm BB} \apgt 3\times10^{4}$ K.  The peak bolometric luminosity is thus $\apgt 2.2\times10^{44}$ ergs s$^{-1}$ and the total energy emitted from integrating under the light-curve model is $\apgt 2.1\times 10^{51}$ ergs, corresponding to a total accreted mass ($M_{\rm acc}$) of $\apgt 0.012 (\epsilon/0.1)^{-1} M_{\odot} $, where $\epsilon$ is the efficiency of converting matter into radiation.  

The internal structure and helium-rich abundance of the star derived from the light curve and the spectra can be consistently modelled by the tidally stripped core of a red giant (precursor to a helium white dwarf) that had a main-sequence mass of $M_{\star} \apgt 1 M_{\odot}$ in order to have evolved off the main sequence in less time than the age of the stellar population ($< 5$ Gyr).  This tidal stripping mechanism has been invoked to explain the hot stars in the Galactic Centre\cite{Davies2005}, and the rate of tidal disruption of tidally stripped stars is likely to be higher than solar-type stars\cite{Kobayashi2004}. 
The mass of the black hole from the light curve fit depends on the mass and radius of the star at the time of disruption.  Using $M_{\star} \sim 0.23 M_{\odot}$ and $R_{\star} \sim 0.33 R_{\odot}$ (measured for a red giant core that was stripped in a binary system\cite{Maxted2011}), and assuming that the evolution of the core is similar to one that is tidally stripped, we find that $f = M_{\rm acc}/M_{\star} \apgt 0.058$ (approaching $f \apgt 0.1$ as measured in simulations\cite{Ayal2000}), that $M_{\rm BH} = (2.8 \pm 0.1) \times 10^{6} M_{\odot}$ and that the peak luminosity approaches the Eddington luminosity of the supermassive black hole ($L_{\rm peak} \apgt 0.6 L_{\rm Edd}$).

\begin{addendum}

\item[Supplementary Information] is linked to the online version of the paper at www.nature.com/nature.

\item[Acknowledgements] We thank H. Tananbaum for approving our \textsl{Chandra} DDT request.  We are grateful to G. Lodato for providing the tidal disruption event mass return-rate models in tabular form, and S. Moran for running software to calculate the host-galaxy K-corrections.  We thank R.E. Williams for helpful discussions on the line emission in the spectra.  S.G. was supported by NASA through a Hubble Fellowship grant awarded by the Space Telescope Science Institute, which is operated by AURA, Inc., for NASA. Partial support for this work was provided by the National Science Foundation.  The Pan-STARRS1 Survey has been made possible through contributions of the Institute for Astronomy, the University of Hawaii, the Pan-STARRS Project Office, the Max-Planck Society and its participating institutes, The Johns Hopkins University, Durham University, the University of Edinburgh, Queen's University Belfast, the Harvard-Smithsonian Center for Astrophysics, and the Las Cumbres Observatory Global Telescope Network, Incorporated, the National Central University of Taiwan, and NASA under a grant issued through the Planetary Science Division of the NASA Science Mission Directorate.
We gratefully acknowledge NASA's support for construction, operation, and science analysis for the GALEX mission, developed in cooperation with Centre National d'Etudes Spatiales of France and the Korean Ministry of Science and Technology. 
Some of the observations reported here were obtained at the MMT Observatory, a
joint facility of the Smithsonian Institution and the University of
Arizona, and the Liverpool Telescope, operated with financial support from the UK Science and Technology Facilities Council.  The computations in this paper were run on the Odyssey cluster supported by the FAS Science Division Research Computing Group at Harvard University.

\item[Contributions] S.G. designed the observations and the transient detection pipeline for \textsl{GALEX} TDS, and measured the UV photometry of PS1-10jh. K.F. and J.D.N coordinated, and D.C.M. facilitated the \textsl{GALEX} TDS observations.  A.R. designed the {\tt photpipe} transient detection pipeline hosted by Harvard/CfA for PS1 MDS, and measured the optical photometry of PS1-10jh.  R.C. designed, implemented, and analysed the MMT optical spectroscopy observations, and contributed to the operation of {\tt photpipe} and the visual inspection of transient alerts.  E.B. proposed for and facilitated the MMT observations.  M.H., G.N., D.S., and R.J.F. contributed to the operation of {\tt photpipe} and the visual inspection of transient alerts.  P.J.C., R.J.F., G.H.M., L.C., and A.S. contributed to the MMT observations.  S.S. designed, and K.S. operated, the transient pipeline for PS1 MDS hosted by Queen's Belfast.  C.W.S., J.L.T., and W.M.W.-V. facilitated the transient pipelines for PS1 MDS.  W.B., K.C.C., T.G., J.N.H., N.K., R.-P.K., E.A.M., J.S.M., P.A.P., C.W.S., and J.L.T. are builders of the PS1 system.  S.G. requested the DDT \textsl{Chandra} X-ray observation and analysed the data.  A.L. obtained the Liverpool Telescope optical imaging observations and analysed the data, and stimulated discussions on the nature of the SED of PS1-10jh.  S.G. analysed and modelled the multi-colour light curve and the SED of PS1-10jh.  T.H. and C.N. stimulated discussions on the nature of the star disrupted.  The paper was put together and written by S.G., and all authors provided feedback on the manuscript.

\item[Author Information] Reprints and permissions information is available at www.nature.com/reprints.  The authors declare no competing financial interests.  Correspondence and requests for materials
should be addressed to S.G.~(email: suvi@pha.jhu.edu).

\end{addendum}


\newpage
\title{\Large\bfseries\noindent SUPPLEMENTARY INFORMATION}


\section{Pan-STARRS1 Medium Deep Survey Observations}

The Pan-STARRS1 system (PS1) is a high-etendue wide-field imaging system designed for dedicated survey observations, on a 1.8~meter telescope 
on Haleakala with a 1.4 Gigapixel camera and a 7 deg$^{2}$ field 
of view\cite{Kaiser2010}.
The PS1 observations are obtained through a set of five broadband
filters, which we have designated as \gps\ ($\lambda_{\rm eff} = 483$ nm), \rps\ ($\lambda_{\rm eff} = 619$ nm), \ips\ ($\lambda_{\rm eff} = 752$ nm), \zps\ ($\lambda_{\rm eff} = 866$ nm), and
\yps ($\lambda_{\rm eff} = 971$ nm).  
Although the filter system for PS1 has much in
common with that used in previous surveys, such as SDSS\cite{Aihara2011}, there
are important differences. The \gps\ filter extends 20~nm redward of
$g_{SDSS}$, paying the price of 5577\AA\ sky emission for greater
sensitivity and lower systematics for photometric redshifts, and the
\zps\ filter is cut off at 930~nm, giving it a different response than
the detector response defined $z_{SDSS}$.  SDSS has no corresponding
\yps\ filter.  

This paper uses images and photometry from the PS1 Medium-Deep Field survey (MDS). The PS1 MDS obtains deep multi-epoch images in the \gps, \rps, \ips, \zps\ and \yps\ bands of 10 fields distributed across the sky chosen for their overlap with extragalactic legacy survey fields with multiwavelength corollary data.  The typical Medium-deep cadence of observations cycles through the \gps, \rps, \ips\, and \zps\ bands every 3 nights, with observations in the \yps\ band close to the full moon.
Images are processed through the Image Processing Pipeline (IPP\cite{Magnier2006}), which runs the images through a succession of stages, 
including flat-fielding (``de-trending''), a flux-conserving warping to a sky-based image plane, masking and artifact removal, and object detection and photometry.  The 8 images taken during any one night are stacked to produce a ``nightly stack''.  This nightly data product is used in two image differencing pipelines which run simultaneously, but independently. 

In this paper, we present photometry from the {\tt photpipe} pipeline hosted at Harvard/CfA\cite{Rest2005}.  This pipeline produces image differences from the nightly stacks and image difference detections which are published to an alerts webpage for visual inspection if there are 3 associated $> 5\sigma$ detections.  Forced-centroid PSF-fitting photometry is applied on its image differences, with a PSF derived from reference stars in each nightly stack.  The zeropoints are measured for the AB system from comparison with field stars in the SDSS catalog.  The photometry is in the natural PS1 system, $m = -2.5 \log(flux) + m'$, with a single zeropoint adjustment $m'$ made in each band to conform to the AB magnitude scale, with an accuracy of better than 1\%.  We do not include the \yps\ band photometry which has an additional uncertainty of $\sim 0.05$ mag in the zeropoint due to the lack of an SDSS comparison.  We propagate the poisson error through the resampling and image differencing. In order to correct for covariance, we do forced photometry in apertures at random positions, calculate the standard deviation of the ratio between the flux and the error, and multiply our errors by this value.  Nightly image differences yield 3$\sigma$ limiting magnitudes of $\sim 23.5$ mag in \gps, \rps, \ips, and \zps\ and a typical positional accuracy of $\sim 0.5$ pixels (0.1 arcsec) which depends on the S/N and FWHM of the source.  The deep template used for the image differencing of PS1-10jh includes the transient flux, and so we also subtract off a negative baseline flux, which is measured from the epochs before the start of the flare in 2009.  We add the error in the mean baseline flux to the photometric error in quadrature.  The image differencing photometry for PS1-10jh is reported in AB magnitudes in Table S1.  In order to improve the signal-to-noise (S/N) in the photometry at late-times ($t > 240$ rest-frame days after the peak) in the figures, we binned the data into time intervals of 30 days.   

\begin{figure}
\caption{
\includegraphics[scale=0.75]{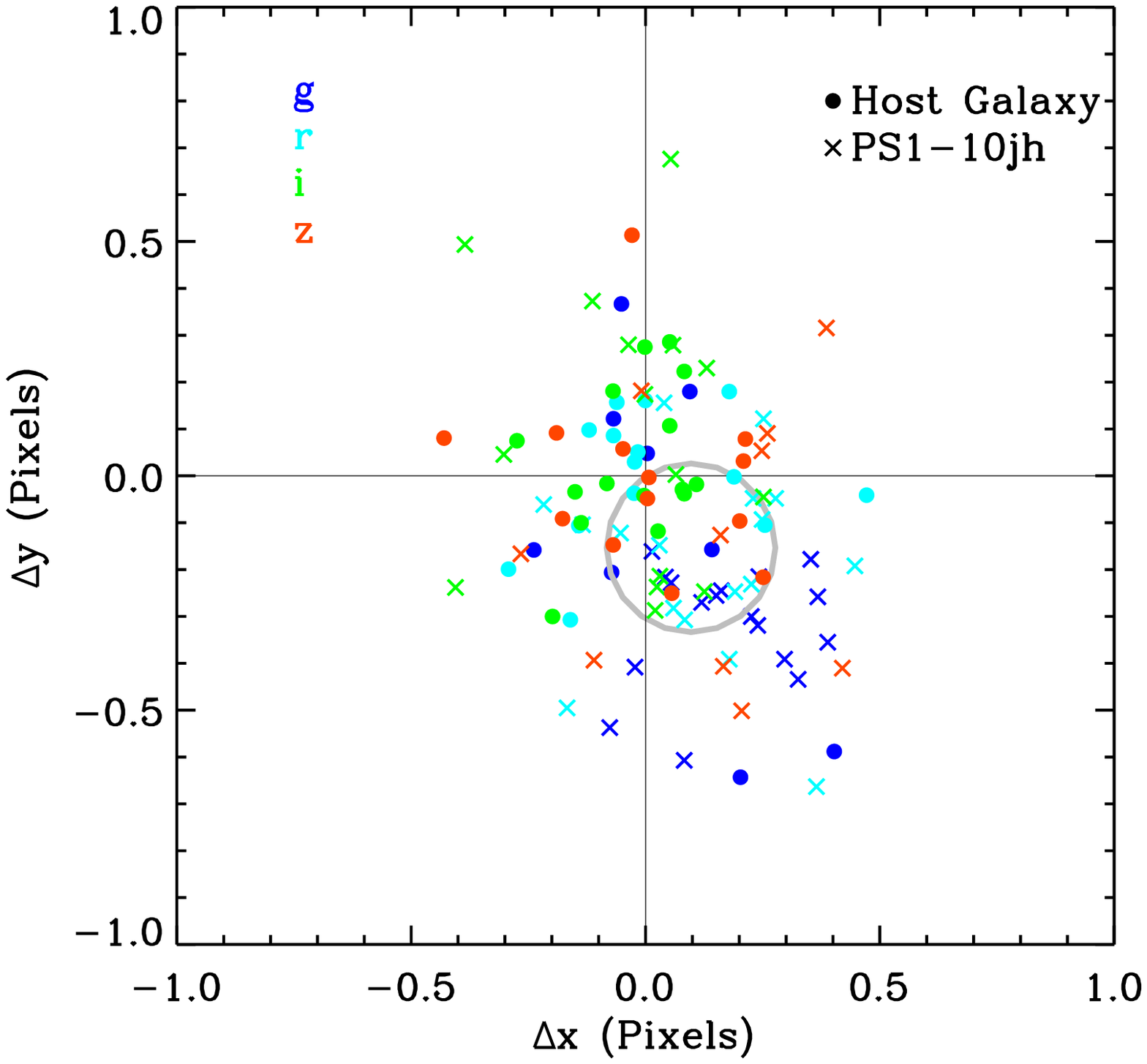}
\\
Offset of PS1-10jh from the mean $x$ and $y$ position of the host galaxy centroid measured in the nightly stacked images before the event.  Solid points show the centroid of the host galaxy before the event, and X symbols show the centroid of PS1-10jh, in each of the 4 PS1 bands.  Thick gray circle shows the mean offset and 3$\sigma$ error of PS1-10jh from the host galaxy centroid.  
\label{fig:pos}
}
\end{figure}

We measure the positional offset between the transient PS1-10jh and the centroid of its host galaxy measured from the nightly stacks before the event.  Figure S\ref{fig:pos} shows the offset in x and y from the mean position of the host galaxy before the event.  The resulting offset during the event is within the 3$\sigma$ uncertainty of 0.18 pixels (0.036 arcsec), plotted with a thick gray circle.  

The PS1 system is developing the Transient Science Server
(TSS) which automatically takes the nightly stacks, 
creates image differences with reference
images created from deep stacks, carries out PSF fitting photometry on the image differences,
and returns catalogues of variable and transient candidates.  
Photometric and astrometric measurements are performed by the
IPP system\cite{Magnier2007, Magnier2008}.  Individual detections made on the
image differences are currently ingested into a MySQL database hosted
at Queen's University Belfast after an initial
culling of objects based on the detection of saturated, masked or
suspected defective pixels within the PSF area. Sources detected on
the nightly image differences are assimilated into potential real
astrophysical transients based on a set of quality tests. 
Transient candidates which pass this automated 
filtering system are promoted for human screening, which currently
runs at around 10\% efficiency (i.e. 10\% of the transients promoted
automatically are judged to be real after human screening).  
Real transients are crossmatched with all available catalogues of 
astronomical sources in the MDS fields (e.g. SDSS, GSC, 2MASS, APM,
Veron AGN,  X-ray catalogues) in order to have a first pass
classification of supernova, variable star, AGN and nuclear transients.

\section{GALEX Time Domain Survey Observations}

\begin{figure}
\caption{
\includegraphics[scale=1.0]{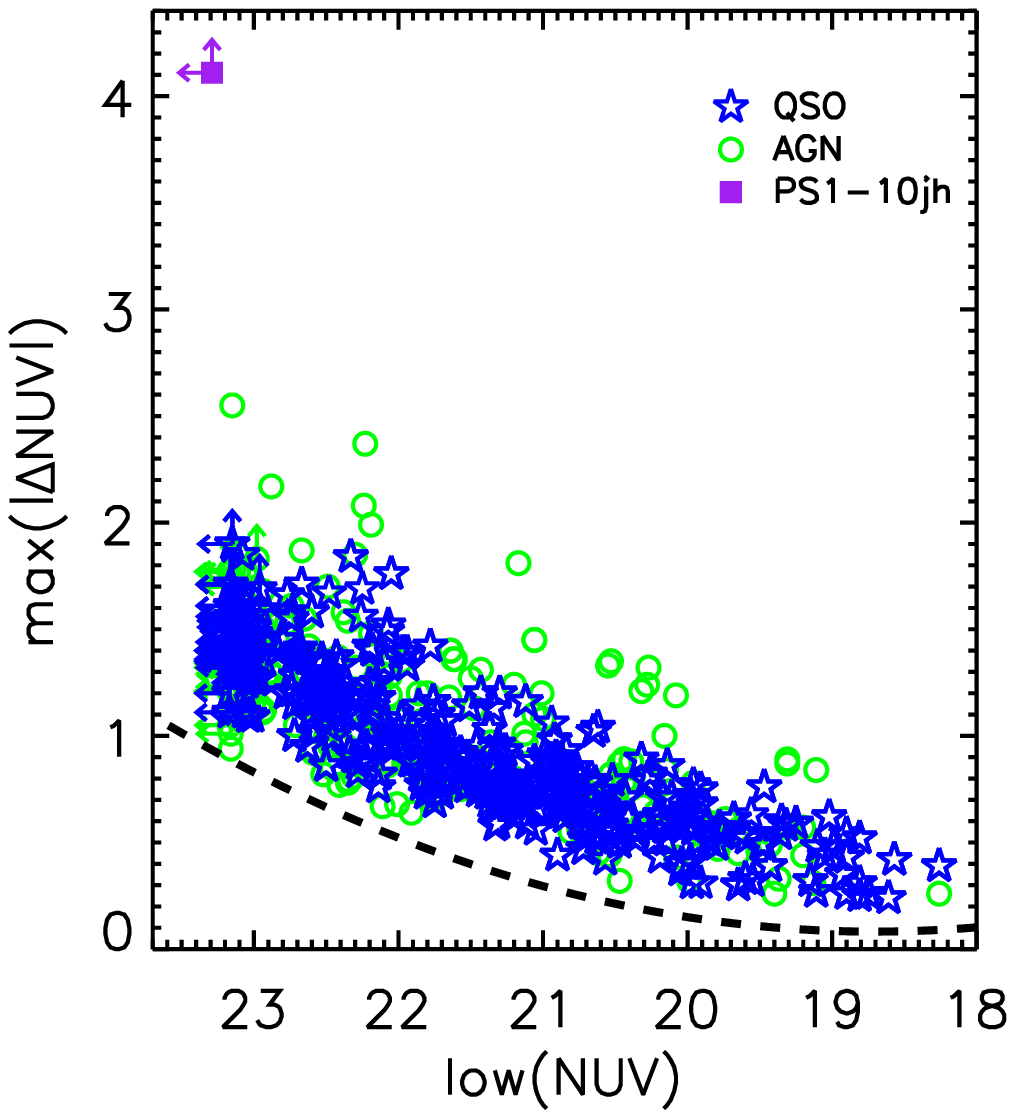}
\\
Maximum amplitude of $NUV$ variability for quasars (blue stars) and active galactic nuclei (green circles) between individual epochs in the \textsl{GALEX} Time Domain Survey.  Dashed line shows the median 5$\sigma$ variability selection function used to select variable source in the \textsl{GALEX} Time Domain Survey fields.  PS1-10jh (purple square) is a clear outlier from these populations, consistent with its $NUV$ flare being a true transient, and not a fluctuation of ongoing accretion activity.  When the pre-event epochs are coadded to a limiting magnitude of $NUV > 25.6$ mag, the peak amplitude of variability of PS1-10jh increases to $> 6.4$ mag.
\label{fig:tds}
}
\end{figure}

The \textsl{GALEX} Time Domain Survey (TDS\cite{Gezari2010}) regularly monitored 6 of the 10 PS1 MDS fields in the \nuv\ ($\lambda_{\rm eff} = 231.6$ nm\cite{Morrissey2007}), with 7 \textsl{GALEX} pointings each with a field of view of $\sim 1$ deg$^{2}$ to cover the full PS1 field of view.  The observations were taken with a cadence of 2 days during the window of observing visibility of each field (from $2-4$ weeks, $1-2$ times per year) from April 2009 to June 2011 UT, and a typical exposure time per epoch of 1.5 ks for a 3$\sigma$ limiting magnitude of $m_{AB} \sim 23.9$ mag.  Variable sources are identified as those which demonstrate an amplitude of variability in any epoch of $>5 \sigma$ from the mean aperture magnitude, where $\sigma$ is determined empirically as a function of magnitude for each epoch from the standard deviation of reference stars in the images. PS1-10jh was discovered independently from PS1 as a transient \nuv source at the 20$\sigma$ level at RA 242.3685 Dec +53.6738 (J2000) on 2010 June 17.68 UT.  The source was undetected in observations between 2009 May 9.52 and 2010 May 9.86 UT.  Figure S\ref{fig:tds} shows the maximum \nuv\ amplitude of UV variable sources classified as quasars and AGNs from the \textsl{GALEX} TDS.  PS1-10jh is a clear outlier, its UV variability is more extreme than variability associated with accretion activity in active galaxies.  The \textsl{GALEX} photometry is measured with a 6 arcsec radius aperture, and corrected for the energy enclosed by the PSF.  The photometry for PS1-10jh is given in AB magnitudes in Table S2.  The 1$\sigma$ error is determined empirically as described above.  To improve the S/N in the photometry at late-times ($t >  240$ rest-frame days after the peak) in the figures, we binned the 8 late-time epochs of data into 3 time intervals in 2011 April, May, and June UT.

\section{MMT Spectroscopy}

\begin{figure}
\caption{
\includegraphics[scale=0.75]{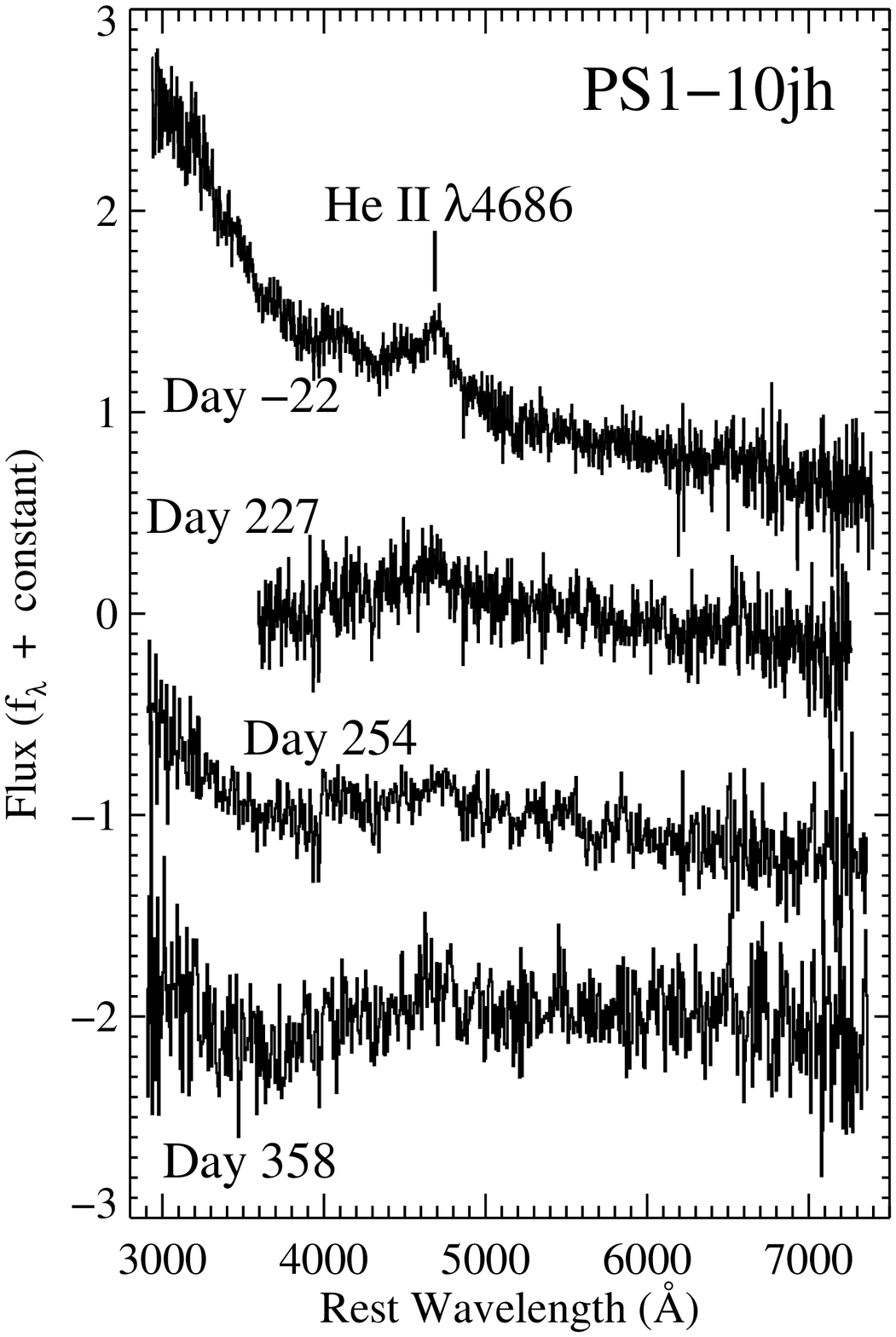}
\\
Series of MMT spectra of PS1-10jh in units of normalized flux density labeled by their phase in rest-frame days since the peak of the flare, plotted with vertical offsets for clarity.  The wavelength of \ion{He}{II}$\lambda 4686$ is labeled with a tick mark.   
\label{fig:spec_all}
}
\end{figure}

We obtained five epochs of optical spectroscopy of PS1-10jh using the
Blue Channel\cite{Schmidt1989} and fiber-fed Hectospec\cite{Fabricant2005}
spectrographs on the 6.5-m MMT.  
We used a long 1 arcsec-wide slit
on the Blue Channel, while the Hectospec fibers are 1.5 arcsec in
diameter. Details of the  observations are presented in
Table~\ref{spectab}.  The Hectospec spectrum was processed using the
standard pipeline\cite{Mink2007} and a flux calibration was applied
using archival observations of the standard star BD$+28\ 4211$.
Basic two-dimensional image processing and extraction of the Blue
Channel data were accomplished using standard routines in IRAF.  We
then used custom IDL routines to apply flux calibrations and remove
telluric absorption based on observations of spectrophotometric
standard stars obtained at similar airmasses.  The absolute flux
scales are unreliable due to clouds and variable seeing on several of
the nights of observations, but the spectra were obtained at the
parallactic angle\cite{Filippenko1982}, so the relative spectral shapes
should be reliable. The effects of second-order light contamination
are apparent in the day 227 Hectospec data at wavelengths $\apgt$
8500~\AA, so we have truncated the spectrum.  We also combined the day
254 and 255 Blue Channel spectra into a single spectrum, and refer to
it as the day 254 spectrum in this paper.  Figure S\ref{fig:spec_all} shows the 
series of spectra.

\begin{figure}
\caption{
\includegraphics[scale=0.75]{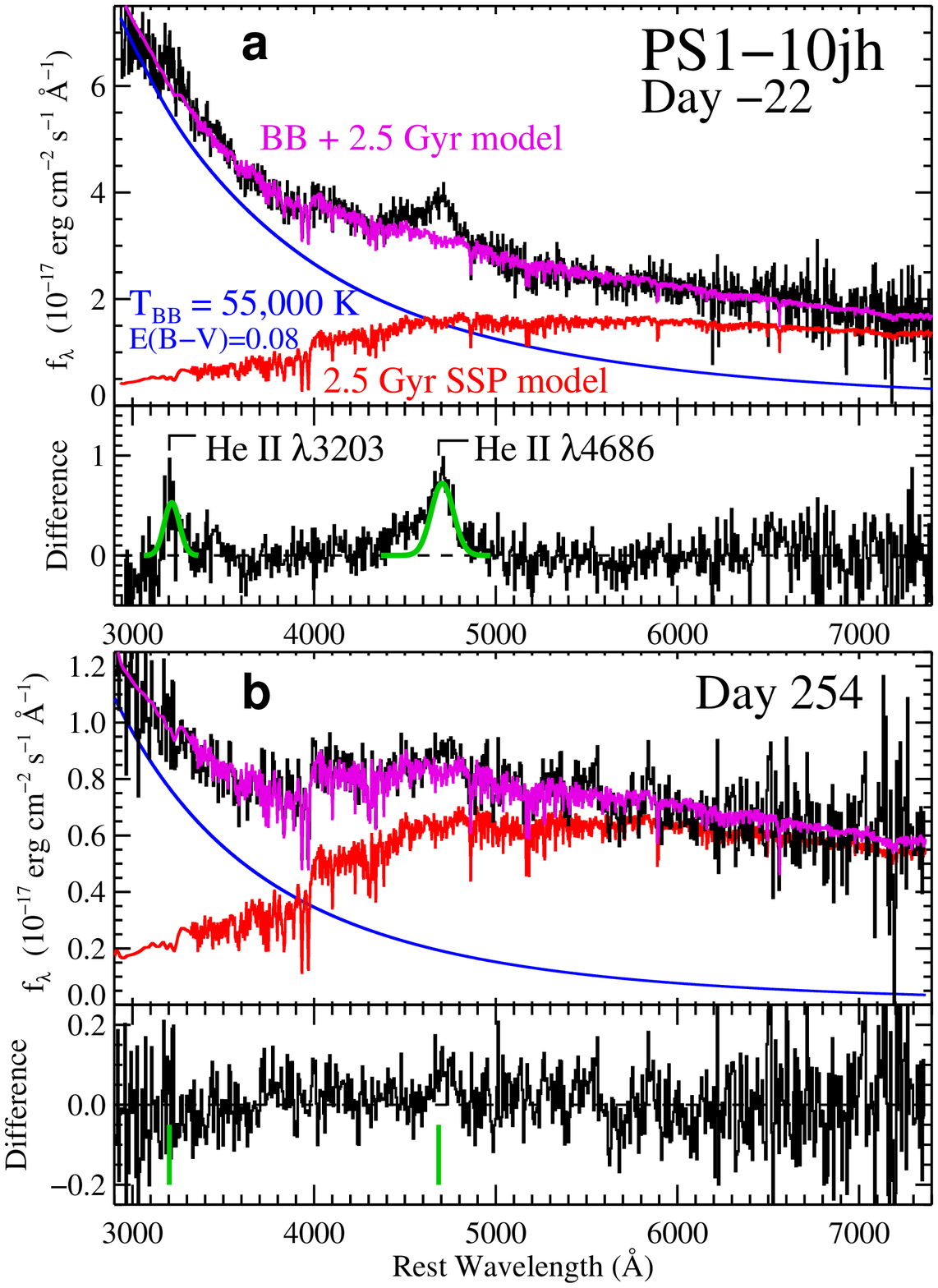}
\\
Same as Figure 1 in the paper, but dereddened for an internal extinction of $E(B-V) = 0.08$ mag, and the continuum fitted with a combination of the same galaxy template and a fading blackbody with $T_{\rm BB} \sim 5.5 \times 10^{4} K$.   
\label{fig:spec_all2}
}
\end{figure}

We created a scaled and weighted stack of all of the post-peak spectra to maximize the S/N in the host spectrum, and fitted the galaxy continuum with template galaxy spectra\cite{Tremonti2004} of different metallicities and stellar populations. The redshift of $z=0.1696 \pm 0.0001$ was determined by cross-correlating with the best-fit templates.  Finally, we performed a chi-squared fit of the models plus a 3$\times 10^{4}$ K blackbody spectrum determined from the UV and optical SED fit, excluding the region around \ion{He}{II}$\lambda$4686.  Simple stellar population (SSP) models with ages in the range $1.4-5$Gyr were all in good agreement with the data.  The formal best fit was found for a 2.5 Gyr SSP with a 1/5th solar metallicity.  However, there is an age-metallicity degeneracy and somewhat younger models at solar metallicity are also a good fit.  The best fit template with a solar metallicity was a 12 Gyr model with an exponentially declining (with an e-folding time of 5 Gyr) star-formation history.  However, none of the results in the paper are sensitive to which exact model is chosen within the set of good matches.  Our spectral resolution (FWHM = 300 km s$^{-1}$) is not sufficient to measure the velocity dispersion ($\sigma_{\star}$) of the host galaxy, which would have $\sigma_{\star} \aplt 100$ km s$^{-1}$ for a central black hole of $< 10^{7} M_{\odot}$.  In Figure S\ref{fig:spec_all2} we show the spectrum dereddened for an internal extinction of $E(B-V) = 0.08$ mag fitted with the same galaxy template as in Figure 2, but with a hotter $5.5\times10^{4}$ K blackbody component.  The quality of the fit is the same with or without internal extinction. 

\section{Chandra Observations}

We requested a 10 ks DDT observation with \textsl{Chandra}\cite{Weisskopf2002} ACIS-S which was obtained on 2011 May 22.96 UT.  No source was detected, with a 3$\sigma$ upper limit of $< 9.4 \times 10^{-4}$ cts s$^{-1}$ calculated using Bayesian statistics with the CIAO v4.3 {\tt aprates} routine for a 4 pixel (1.968 arcsec) radius aperture.  This corresponds to a flux of $< 7.2 \times 10^{-15}$ ergs s$^{-1}$ cm$^{-2}$ when corrected for Galactic extinction with $N_{H} = 3.1 E(B-V) 1.8\times10^{21}$ cm$^{-2} = 7.2 \times 10^{19}$ cm$^{-2}$ and assuming a $\Gamma = 2$ energy spectrum typical of an unobscured AGN, or $L_{X} (0.2-10) $keV$ < 5.8 \times 10^{41}$ ergs s$^{-1}$.  The upper limit to the $\alpha_{ox}$ ratio using the $NUV$ observation closest in time on 2011 May 12.37 UT with $L_{\nu} = (4.3 \pm 0.7) \times 10^{27}$ ergs s$^{-1}$ Hz$^{-1}$ (corrected for Galactic extinction) is $\alpha_{\rm ox} = \log [L_{\nu}(2500 $\AA$)/L_{\nu}(2 $keV$)]/\log[\nu(2500 $\AA$)/\nu(2 $keV$)] < (-1.65 \pm 0.03)$, well below the mean for broad-lined AGN of comparable \nuv\ luminosity of $\alpha_{\rm ox} \sim -1.15$ (Steffen \etal 2006).

The non-detection by \textsl{Chandra} is consistent with blackbody emission of $\aplt 2.5 \times 10^{5}$ K for bolometric luminosities of up to $\sim 10^{44}$ ergs s$^{-1}$, close to the Eddington luminosity of the central black hole.  For higher blackbody temperatures, such as the range observed in the X-ray TDE candidates from \textsl{ROSAT}, \textsl{XMM-Newton}, and \textsl{Chandra}\cite{Gezari2009} of $6-12 \times 10^{5}$ K, the \textsl{Chandra} non-detection of PS1-10jh places an upper limit on the bolometric luminosity of such a blackbody component of $\sim 10^{42}$ ergs s$^{-1}$, below the luminosities of $10^{42-44}$ ergs s$^{-1}$ of the X-ray TDE candidates.  However, the blackbody temperatures of the X-ray TDE candidates are hotter than expected for a TDE from basic theoretical arguments, and correspond to effective radii smaller than the Schwarzschild radius of their respective black holes.  A lower effective temperature of $\aplt 2.5 \times 10^{5}$ K, and thus a non-detection in the hard X-rays, is actually in better agreement with theoretical expectations for thermal emission from radii ranging from the innermost stable circular orbit $(R_{\rm ISCO})$ to the tidal disruption radius of the central black hole ($R_{\rm T}$).  Furthermore, it would not be surprising if TDE candidates selected from X-ray surveys were more X-ray bright than those selected using other methods.

\section{Nature of the Flare}

The persistence of the hot blackbody emission up to 375 rest-frame days after the peak definitively excludes a supernova (SN) origin.  Although core-collapse SNe are hot at early times ($\sim 10^{4}$ K), they quickly cool through expansion and radiation to $\sim 6000$ K by a month after explosion (i.e., Type II SNe\cite{Hamuy1988, Dessart2008, Gezari2010}, Type Ibc SNe\cite{Soderberg2008}, Ultraluminous SNe\cite{Gezari2009b, Chomiuk2011}).  The lack of recent star-formation in the host galaxy also disfavors a core-collapse SN with a massive progenitor star with a short lifetime.  The host galaxy is undetected in a deep coadd of all the \textsl{GALEX} TDS epochs in 2009 in the $FUV (\lambda_{\rm eff} = 153.9$ nm) with $t_{\rm exp} = 14.9$ ks and $NUV$ with $t_{\rm exp} = 43.2$ ks, with 3$\sigma$ upper limits of $FUV > 25.1$ mag and $NUV > 25.6$ mag.  The upper limit on the $NUV$ flux density corresponds to an upper limit on the star-formation rate\cite{Kennicutt1998} in the host galaxy of $< 0.022 M_{\odot}$ yr$^{-1}$ after correcting for Galactic extinction.

The upper limit to the X-ray to UV luminosity density ratio $260-270$ rest-frame days from the peak is 20 times lower than observed in broad-lined AGNs of a comparable \nuv\ luminosity\cite{Steffen2006}, and argues strongly against an association of the flare with an AGN.  Furthermore, the extreme amplitude of the flare of $ > 6.4$ mag is most likely caused by a true transient event, and not from a fluctuation of unobscured accretion activity.

The amplitude of the flare could be explained by a change in the line-of-sight extinction toward the nucleus of the galaxy of $\Delta(N_{H}) = 5 \times 10^{21}$ cm$^{-2}$.  However, in order to obscure the AGN hard X-ray emission during the flare, assuming a standard intrinsic $\alpha_{\rm ox}$, one requires $N_{H} \sim 10^{24}$ cm$^{-2}$.  With such a high column density, for a standard gas-to-dust ratio the UV and optical extinction would be extremely large ($E(B-V) \sim 180$ mag), and no UV and optical flare from the nucleus would be observable.  

X-ray bright optically normal galaxies (XBONGs) have been observed which show strong X-ray emission characteristic of an unobscured AGN, but with no optical emission lines characteristic of AGN activity.  This scenario has been explained by an AGN with a unobscured nucleus whose optical nuclear emission lines are diluted by a strong stellar continuum\cite{Georg2005}.  These sources share the property with PS1-10jh in the lack of a standard AGN emission-line spectrum, however the detection of broad He~II emission in PS1-10jh indicates that its optical nuclear spectrum is neither diluted nor absorbed.


\section{Light Curve Fits to Tidal Disruption Accretion Rate Models}

The index $n$ of the power-law decay, $\dot{M} \propto (t/t_{\rm min})^{-n}$, is sensitive to mode of the accretion.  For super-Eddington accretion rates, a radiation supported outflow expands with a receding photosphere\cite{Strubbe2009}, resulting in a brief outburst that peaks at $\sim 10 (M_{\rm BH}/10^{6} M_{\odot})^{-1/8}(R_{\rm T}/R_{\rm p})^{-9/8}m_{\star}r_{\star}^{6/8}$ d, and then declines in luminosity as $n=5/9$.  Emission on the Rayleigh-Jeans tail ($L_{\nu} \propto T$) of the hot outflow declines as $n=35/36$.  This power-law index can be fitted to the decay of PS1-10jh, but with $t_{\rm min} = 9 \pm 3$ rest-frame days, which is incompatible with the observed rise-time of the flare of $> 35$ rest-frame days. 

For sub-Eddington accretion rates, the luminosity should follow the decline of the mass-return rate, which depends on the internal structure of the star at early times, but approaches an $n=5/3$ power-law after a few times $t_{\rm min}$ for all stellar types.  For $L = 4\pi R_{\rm BB}^{2}\sigma T_{\rm BB}^{4}$, if $L \propto \dot{M} \propto (t/t_{\rm min})^{-5/3}$, then on the Rayleigh-Jeans tail, for a fixed $R_{\rm BB}$ one expects an $n=5/12$ decay\cite{Strubbe2009, Lodato2011}.  Our $NUV$ and optical photometry of PS1-10jh are on the Rayleigh-Jeans tail of the $\apgt 3 \times 10^{4}$ K blackbody, yet we see a decline that follows the mass-return rate, with no indication of a shallower decline due to cooling of the emission.  We also do not observe any evolution in the UV and optical SED that would indicate cooling over time.  

A possible explanation for both the constant shape of the UV and optical SED and the linear scaling of the UV and optical light curve with the predicted bolometric luminosity evolution of the TDE, is that the UV and optical continuum is a "pseudo-continuum" whose shape is determined by atomic reprocessing.  In such a scenario, the UV and optical SED shape remains fixed even if the photoionising continuum is cooling with time (its shape is determined by a velocity-blurred reflection spectrum and not the temperature of the photoionising continuum), and the UV and optical light follows the decay of the bolometric luminosity since it is the result of the reflection, absorption, and re-emission of the photoionising continuum.  This explanation implies that the expected very hot $\approx 10^{5}$ K blackbody photoionising continuum is present in the unseen EUV region, but is masked in the observed UV and optical region by reprocessing. This model has also been invoked for normal AGNs to explain their low apparent thermal UV and optical continuum temperature (the Big Blue Bump)\cite{Lawrence2011}.  

\begin{figure}
\caption{
\includegraphics[scale=0.75]{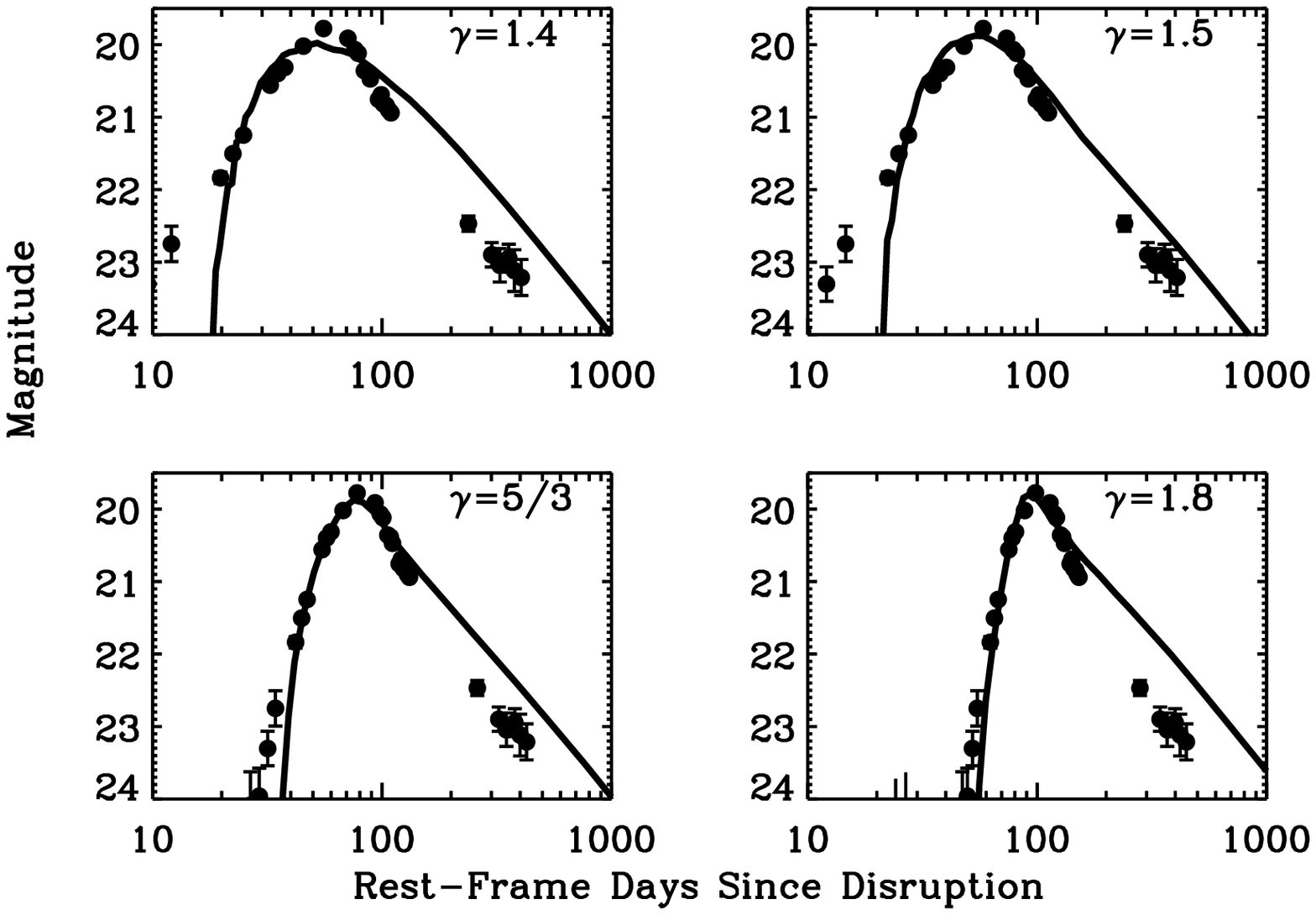}
\\
Fits of the \gps-band light curve of PS1-10jh from $-38$ to $58$ rest-frame days from the peak to models for the mass accretion rate of tidally disrupted stars of different polytropic exponent $\gamma$. 
\label{fig:rise_models}
}
\end{figure}

The rise and decay of PS1-10jh is well constrained by the PS1 photometry, and enables us to determine the polytropic exponent ($\gamma$) of the star disrupted.  Figure S\ref{fig:rise_models} shows the fit to the \gps\ light curve for models\cite{Lodato2009} with $\gamma = 1.4, 1.5, 5/3,$ and $1.8$.  The data are in best agreement with the $\gamma = 5/3$ model.  The derived parameters from the fits are stretch factors in time of $1.25, 1.09, 1.40$, and $1.77$, respectively, and a time delay between the time of disruption and the peak of the flare of $56, 58, 78,$ and $98$ rest-frame days, respectively.  Without the constraints from the rise {\it and} decay of the light curve, the values for the stretch factor and the time of disruption can vary widely.

\section{He Abundance and Internal Extinction}

The He/H ratio is derived from $f($\ion{He}{II}$\lambda4686$)$/f$(H$\alpha$) = $\frac{n({\rm He}^{+})\alpha^{eff}_{\lambda 4686}h\nu_{\lambda 4686}}{n({\rm H}^{0})\alpha^{eff}_{\rm H\beta}(j_{\rm H\alpha}/j_{\rm H\beta})h\nu_{\rm H\beta}}$, where $j_{\rm H\alpha}/j_{\rm H\beta} = 3.1$ includes the effects of collisional excitation, $\alpha^{eff}_{H\beta} = 3.03\times10^{-14}$ cm$^{3}$ s$^{-1}$ and $\alpha^{eff}_{\lambda 4686} = 3.72\times10^{-13}$ cm$^{3}$ s$^{-1}$ for a gas temperature of $T = 1 \times 10^{4}$ K typical of nebular gas and the broad-line region of an AGN. The 3$\sigma$ upper limit measured from the noise in the continuum of H$\alpha/$\ion{He}{II}$\lambda 4686 < 0.2$ implies a He abundance of $n($He$^{+})/n($H$^{0}) > 1.2$, which corresponds to a hydrogen mass fraction of $X = \frac{n_{H}}{n_{H} + 4n_{He}} < 0.2$. Since the number density of the unbound debris is high\cite{Strubbe2009}, $n \sim 3 \times 10^{13} M_{6}^{1/6} \beta^{-5} m_{\star}^{-2/3} r_{\star}^{3/2} (t/36$ d$)^{-3}$ cm$^{-3}$, the recombination time is short compared to the flare timescale, $\tau_{rec} = (n_{e}\alpha_{B})^{-1} \sim (n\frac{1+2[n({\rm H}^{0})/n({\rm He}^{+})]}{1+n({\rm H}^{0})/n({\rm He}^{+})}\alpha_{B})^{-1} \approx 0.08 (\frac{t}{36 {\rm d}})^{3}$ sec, and we can assume that the gas reaches photoionization equilibrium instantaneously.
We derive the internal extinction from $E(B-V)_{\rm int} = \frac{\log (R_{\rm obs}/R_{\rm int})}{-0.4[k(\lambda 3203)-k(\lambda 4686)]} < 0.08$ mag, where $R_{\rm obs}$ and $R_{\rm int}$ are the observed and intrinsic \ion{He}{II} $\lambda 3203/\lambda 4686$ ratios, and we use $R_{\rm obs} = 0.5 \pm 0.1$ and $R_{\rm int} = 0.45$, and an extinction law\cite{Cardelli1989} with $k(\lambda 3203)-k(\lambda 4686) = 1.555$.

\clearpage

\begin{longtable}{llrrcrr}
\caption{Pan-STARRS1 Medium Deep Survey Observations \label{ps1tab}}\\
\hline
\hline
\multicolumn{3}{c}{UT Date} & Phase & Filter & Mag & $\sigma$\\
\hline
2009 & May & 28.48 & -350.40 & \gps & $>$23.90 &  \\
2009 & May & 31.43 & -347.88 & \gps & $>$23.89 &  \\
2009 & Jun & 14.45 & -335.90 & \gps & $>$23.70 &  \\
2009 & Jun & 17.44 & -333.34 & \gps & $>$23.79 &  \\
2009 & Jun & 20.42 & -330.79 & \gps & $>$23.83 &  \\
2009 & Jul &  2.29 & -320.64 & \gps & $>$23.62 &  \\
2009 & Jul & 17.30 & -307.80 & \gps & $>$23.04 &  \\
2009 & Sep & 15.25 & -256.55 & \gps & $>$23.84 &  \\
2010 & Apr & 16.59 &  -74.15 & \gps & $>$23.72 &  \\
2010 & Apr & 19.58 &  -71.59 & \gps & $>$23.63 &  \\
2010 & May & 10.55 &  -53.66 & \gps & 25.43 & 1.49 \\
2010 & May & 13.53 &  -51.11 & \gps & 24.09 & 0.46 \\
2010 & May & 16.43 &  -48.64 & \gps & 23.96 & 0.39 \\
2010 & May & 19.43 &  -46.07 & \gps & 23.30 & 0.24 \\
2010 & May & 22.43 &  -43.50 & \gps & 22.75 & 0.24 \\
2010 & May & 31.45 &  -35.80 & \gps & 21.84 & 0.08 \\
2010 & Jun &  3.51 &  -33.17 & \gps & 21.50 & 0.05 \\
2010 & Jun &  6.42 &  -30.69 & \gps & 21.25 & 0.03 \\
2010 & Jun & 15.37 &  -23.04 & \gps & 20.56 & 0.02 \\
2010 & Jun & 18.33 &  -20.50 & \gps & 20.40 & 0.02 \\
2010 & Jun & 21.40 &  -17.88 & \gps & 20.31 & 0.02 \\
2010 & Jun & 30.44 &  -10.15 & \gps & 20.02 & 0.01 \\
2010 & Jul & 12.31 &    0.00 & \gps & 19.78 & 0.01 \\
2010 & Jul & 30.34 &   15.41 & \gps & 19.91 & 0.01 \\
2010 & Aug &  5.32 &   20.53 & \gps & 20.07 & 0.01 \\
2010 & Aug &  8.30 &   23.07 & \gps & 20.12 & 0.01 \\
2010 & Aug & 14.29 &   28.19 & \gps & 20.36 & 0.02 \\
2010 & Aug & 17.26 &   30.73 & \gps & 20.38 & 0.02 \\
2010 & Aug & 20.27 &   33.30 & \gps & 20.47 & 0.02 \\
2010 & Aug & 29.26 &   41.00 & \gps & 20.75 & 0.02 \\
2010 & Sep &  1.29 &   43.58 & \gps & 20.69 & 0.02 \\
2010 & Sep &  4.26 &   46.12 & \gps & 20.82 & 0.02 \\
2010 & Sep &  7.27 &   48.70 & \gps & 20.84 & 0.02 \\
2010 & Sep & 10.23 &   51.23 & \gps & 20.90 & 0.03 \\
2010 & Sep & 13.25 &   53.81 & \gps & 20.94 & 0.03 \\
2011 & Feb & 10.62 &  182.37 & \gps & 22.47 & 0.10 \\
2011 & Apr & 23.51 &  243.84 & \gps & 23.03 & 0.20 \\
2011 & Apr & 26.58 &  246.47 & \gps & 22.77 & 0.13 \\
2011 & May & 20.57 &  266.97 & \gps & 23.10 & 0.29 \\
2011 & May & 29.46 &  274.58 & \gps & 22.99 & 0.16 \\
2011 & Jun & 10.38 &  284.77 & \gps & $>$22.79 &  \\
2011 & Jun & 25.39 &  297.61 & \gps & 22.75 & 0.14 \\
2011 & Jun & 28.38 &  300.16 & \gps & 23.09 & 0.28 \\
2011 & Jul &  1.32 &  302.68 & \gps & 23.20 & 0.19 \\
2011 & Jul &  4.33 &  305.25 & \gps & 22.74 & 0.12 \\
2011 & Jul & 10.31 &  310.36 & \gps & 22.74 & 0.19 \\
2011 & Jul & 19.28 &  318.03 & \gps & 23.28 & 0.21 \\
2011 & Jul & 22.28 &  320.60 & \gps & 23.54 & 0.38 \\
2011 & Jul & 25.28 &  323.16 & \gps & 22.96 & 0.16 \\
2011 & Jul & 28.28 &  325.72 & \gps & 22.49 & 0.10 \\
2011 & Jul & 31.28 &  328.29 & \gps & 22.81 & 0.14 \\
2011 & Aug &  3.30 &  330.87 & \gps & 23.04 & 0.16 \\
2011 & Aug &  6.27 &  333.41 & \gps & 24.08 & 0.60 \\
2011 & Aug & 18.29 &  343.69 & \gps & 22.96 & 0.15 \\
2011 & Aug & 21.27 &  346.23 & \gps & 23.07 & 0.16 \\
2011 & Aug & 24.27 &  348.80 & \gps & 23.14 & 0.18 \\
2011 & Aug & 27.25 &  351.35 & \gps & 22.87 & 0.20 \\
2011 & Aug & 30.28 &  353.94 & \gps & 24.01 & 0.43 \\
\hline
2009 & Apr & 20.62 & -382.77 & \rps & $>$23.35 &  \\
2009 & Apr & 29.61 & -375.09 & \rps & $>$23.30 &  \\
2009 & Apr & 30.60 & -374.24 & \rps & $>$23.34 &  \\
2009 & May &  2.59 & -372.54 & \rps & $>$23.32 &  \\
2009 & May & 22.52 & -355.50 & \rps & $>$23.37 &  \\
2009 & Jun & 11.46 & -338.45 & \rps & $>$23.26 &  \\
2009 & Jun & 14.46 & -335.89 & \rps & $>$23.20 &  \\
2009 & Jun & 17.46 & -333.32 & \rps & $>$23.06 &  \\
2009 & Jun & 20.43 & -330.78 & \rps & $>$23.31 &  \\
2009 & Jul &  2.31 & -320.63 & \rps & $>$23.29 &  \\
2009 & Sep & 15.26 & -256.54 & \rps & $>$23.32 &  \\
2010 & Apr & 19.60 &  -71.58 & \rps & $>$23.20 &  \\
2010 & May & 16.44 &  -48.63 & \rps & 24.21 & 0.81 \\
2010 & May & 19.45 &  -46.06 & \rps & 23.34 & 0.39 \\
2010 & May & 22.45 &  -43.49 & \rps & 22.71 & 0.29 \\
2010 & May & 31.46 &  -35.78 & \rps & 22.46 & 0.17 \\
2010 & Jun &  3.53 &  -33.16 & \rps & 21.80 & 0.09 \\
2010 & Jun &  6.44 &  -30.67 & \rps & 21.50 & 0.07 \\
2010 & Jun & 15.38 &  -23.03 & \rps & 20.94 & 0.04 \\
2010 & Jun & 18.32 &  -20.51 & \rps & 20.69 & 0.03 \\
2010 & Jun & 21.41 &  -17.87 & \rps & 20.63 & 0.03 \\
2010 & Jun & 30.45 &  -10.14 & \rps & 20.32 & 0.03 \\
2010 & Jul & 12.30 &   -0.01 & \rps & 20.12 & 0.02 \\
2010 & Jul & 30.35 &   15.42 & \rps & 20.23 & 0.02 \\
2010 & Aug &  5.31 &   20.52 & \rps & 20.36 & 0.02 \\
2010 & Aug &  8.32 &   23.09 & \rps & 20.46 & 0.03 \\
2010 & Aug & 14.30 &   28.21 & \rps & 20.69 & 0.05 \\
2010 & Aug & 17.27 &   30.75 & \rps & 20.70 & 0.04 \\
2010 & Aug & 20.28 &   33.31 & \rps & 20.72 & 0.04 \\
2010 & Aug & 29.27 &   41.01 & \rps & 21.08 & 0.05 \\
2010 & Sep &  1.27 &   43.57 & \rps & 21.10 & 0.05 \\
2010 & Sep &  4.27 &   46.13 & \rps & 21.15 & 0.05 \\
2010 & Sep &  7.29 &   48.71 & \rps & 21.18 & 0.05 \\
2010 & Sep & 10.24 &   51.24 & \rps & 21.22 & 0.05 \\
2010 & Sep & 13.27 &   53.82 & \rps & 21.38 & 0.07 \\
2011 & Feb & 10.63 &  182.38 & \rps & 22.47 & 0.17 \\
2011 & Apr & 23.52 &  243.85 & \rps & 23.01 & 0.27 \\
2011 & Apr & 26.59 &  246.48 & \rps & 23.58 & 0.46 \\
2011 & May & 20.58 &  266.98 & \rps & 23.41 & 0.47 \\
2011 & May & 29.48 &  274.59 & \rps & 23.57 & 0.45 \\
2011 & Jun & 10.39 &  284.78 & \rps & 22.78 & 0.27 \\
2011 & Jun & 25.41 &  297.62 & \rps & 23.04 & 0.31 \\
2011 & Jun & 28.39 &  300.17 & \rps & 24.37 & 1.52 \\
2011 & Jul &  1.34 &  302.69 & \rps & 23.48 & 0.43 \\
2011 & Jul &  4.34 &  305.26 & \rps & 23.25 & 0.33 \\
2011 & Jul & 10.32 &  310.37 & \rps & 24.01 & 0.76 \\
2011 & Jul & 19.29 &  318.04 & \rps & 23.01 & 0.28 \\
2011 & Jul & 22.29 &  320.61 & \rps & 23.39 & 0.54 \\
2011 & Jul & 25.30 &  323.17 & \rps & 23.60 & 0.46 \\
2011 & Jul & 28.29 &  325.74 & \rps & 23.10 & 0.29 \\
2011 & Jul & 31.29 &  328.30 & \rps & 23.44 & 0.42 \\
2011 & Aug &  3.32 &  330.89 & \rps & 23.53 & 0.43 \\
2011 & Aug &  6.28 &  333.42 & \rps & 24.59 & 1.24 \\
2011 & Aug & 18.30 &  343.70 & \rps & 23.89 & 0.60 \\
2011 & Aug & 21.28 &  346.25 & \rps & 23.41 & 0.38 \\
2011 & Aug & 24.28 &  348.81 & \rps & 23.46 & 0.40 \\
2011 & Aug & 30.29 &  353.95 & \rps & 24.19 & 0.80 \\
\hline
2009 & Apr & 19.55 & -383.69 & \ips & $>$22.65 &  \\
2009 & Apr & 20.59 & -382.80 & \ips & $>$22.93 &  \\
2009 & May &  1.61 & -373.38 & \ips & $>$22.90 &  \\
2009 & May &  2.55 & -372.57 & \ips & $>$22.84 &  \\
2009 & Jun &  2.42 & -346.18 & \ips & $>$22.94 &  \\
2009 & Jun &  3.47 & -345.28 & \ips & $>$22.91 &  \\
2009 & Jun & 15.37 & -335.11 & \ips & $>$22.97 &  \\
2009 & Jun & 18.40 & -332.52 & \ips & $>$22.93 &  \\
2009 & Jun & 30.36 & -322.29 & \ips & $>$22.95 &  \\
2009 & Jul &  3.32 & -319.76 & \ips & $>$22.84 &  \\
2009 & Sep &  1.25 & -268.52 & \ips & $>$22.86 &  \\
2009 & Sep & 19.24 & -253.14 & \ips & $>$22.88 &  \\
2010 & Apr &  2.54 &  -86.16 & \ips & $>$22.91 &  \\
2010 & Apr & 14.59 &  -75.85 & \ips & $>$22.93 &  \\
2010 & May &  8.59 &  -55.34 & \ips & $>$22.94 &  \\
2010 & May & 11.61 &  -52.76 & \ips & 24.45 & 1.44 \\
2010 & Jun &  1.41 &  -34.97 & \ips & 22.59 & 0.26 \\
2010 & Jun & 16.50 &  -22.07 & \ips & 21.00 & 0.06 \\
2010 & Jun & 19.33 &  -19.65 & \ips & 20.85 & 0.05 \\
2010 & Jul &  1.35 &   -9.37 & \ips & 20.48 & 0.04 \\
2010 & Jul & 31.30 &   16.23 & \ips & 20.35 & 0.03 \\
2010 & Aug &  3.35 &   18.84 & \ips & 20.47 & 0.04 \\
2010 & Aug &  6.33 &   21.39 & \ips & 20.51 & 0.04 \\
2010 & Aug &  9.28 &   23.91 & \ips & 20.52 & 0.04 \\
2010 & Aug & 15.28 &   29.04 & \ips & 20.69 & 0.04 \\
2010 & Aug & 30.29 &   41.88 & \ips & 21.27 & 0.08 \\
2010 & Sep &  2.27 &   44.42 & \ips & 21.14 & 0.07 \\
2010 & Sep &  5.27 &   46.98 & \ips & 21.24 & 0.07 \\
2010 & Sep &  8.26 &   49.55 & \ips & 21.31 & 0.08 \\
2010 & Sep & 11.26 &   52.11 & \ips & 21.43 & 0.09 \\
2010 & Sep & 14.25 &   54.66 & \ips & 21.47 & 0.09 \\
2010 & Sep & 17.23 &   57.21 & \ips & 21.52 & 0.10 \\
2011 & Feb & 23.63 &  193.50 & \ips & 23.45 & 0.83 \\
2011 & Apr & 21.51 &  242.13 & \ips & 23.05 & 0.40 \\
2011 & Apr & 24.58 &  244.76 & \ips & 23.18 & 0.45 \\
2011 & Apr & 27.55 &  247.30 & \ips & 23.72 & 0.74 \\
2011 & May & 12.43 &  260.01 & \ips & 23.99 & 1.57 \\
2011 & May & 21.52 &  267.79 & \ips & 23.88 & 0.88 \\
2011 & May & 27.43 &  272.84 & \ips & 24.07 & 1.04 \\
2011 & May & 30.49 &  275.46 & \ips & 23.67 & 0.70 \\
2011 & Jun &  2.40 &  277.95 & \ips & 23.92 & 0.87 \\
2011 & Jun & 11.35 &  285.60 & \ips & 23.18 & 0.45 \\
2011 & Jun & 26.39 &  298.46 & \ips & 23.12 & 0.49 \\
2011 & Jun & 29.46 &  301.08 & \ips & 23.66 & 0.70 \\
2011 & Jul &  2.32 &  303.53 & \ips & 23.44 & 0.57 \\
2011 & Jul &  5.34 &  306.11 & \ips & 23.47 & 0.58 \\
2011 & Jul & 11.29 &  311.20 & \ips & 23.60 & 0.65 \\
2011 & Jul & 20.34 &  318.93 & \ips & 24.10 & 1.04 \\
2011 & Jul & 23.29 &  321.45 & \ips & 23.47 & 0.62 \\
2011 & Jul & 26.29 &  324.02 & \ips & 23.82 & 0.79 \\
2011 & Jul & 29.32 &  326.61 & \ips & 23.74 & 0.78 \\
2011 & Aug &  1.27 &  329.14 & \ips & 23.39 & 0.53 \\
2011 & Aug &  4.27 &  331.70 & \ips & 23.38 & 0.53 \\
2011 & Aug & 16.28 &  341.97 & \ips & 23.67 & 0.71 \\
2011 & Aug & 19.31 &  344.56 & \ips & 23.22 & 0.47 \\
2011 & Aug & 22.27 &  347.09 & \ips & 23.67 & 0.70 \\
2011 & Aug & 28.26 &  352.21 & \ips & 24.11 & 1.04 \\
2011 & Aug & 31.27 &  354.79 & \ips & 23.13 & 0.43 \\
2011 & Sep &  3.25 &  357.33 & \ips & 24.06 & 1.04 \\
\hline
2009 & May &  5.46 & -370.08 & \zps & $>$22.97 &  \\
2009 & May &  6.55 & -369.15 & \zps & $>$22.82 &  \\
2009 & Jun & 13.42 & -336.78 & \zps & $>$23.04 &  \\
2009 & Jun & 16.43 & -334.20 & \zps & $>$22.88 &  \\
2009 & Jun & 25.39 & -326.54 & \zps & $>$22.97 &  \\
2009 & Jun & 28.37 & -324.00 & \zps & $>$23.02 &  \\
2009 & Jul &  4.30 & -318.93 & \zps & $>$22.97 &  \\
2009 & Jul & 22.39 & -303.45 & \zps & $>$22.64 &  \\
2009 & Sep & 20.24 & -252.29 & \zps & $>$22.19 &  \\
2009 & Sep & 29.22 & -244.60 & \zps & $>$23.02 &  \\
2010 & Apr &  9.62 &  -80.11 & \zps & $>$22.90 &  \\
2010 & Apr & 12.61 &  -77.55 & \zps & $>$23.11 &  \\
2010 & Apr & 18.59 &  -72.44 & \zps & $>$23.06 &  \\
2010 & May &  9.52 &  -54.55 & \zps & $>$22.96 &  \\
2010 & May & 21.51 &  -44.29 & \zps & 24.88 & 2.03 \\
2010 & Jun &  5.46 &  -31.50 & \zps & 21.82 & 0.11 \\
2010 & Jun & 11.55 &  -26.30 & \zps & 21.51 & 0.10 \\
2010 & Jun & 14.37 &  -23.89 & \zps & 21.28 & 0.07 \\
2010 & Jun & 17.40 &  -21.30 & \zps & 21.47 & 0.10 \\
2010 & Jun & 20.31 &  -18.81 & \zps & 20.97 & 0.05 \\
2010 & Jun & 29.39 &  -11.04 & \zps & 20.64 & 0.04 \\
2010 & Jul &  2.29 &   -8.57 & \zps & 20.67 & 0.04 \\
2010 & Jul & 29.39 &   14.61 & \zps & 20.62 & 0.04 \\
2010 & Aug &  1.34 &   17.13 & \zps & 20.57 & 0.04 \\
2010 & Aug &  4.38 &   19.72 & \zps & 20.64 & 0.04 \\
2010 & Aug & 16.30 &   29.92 & \zps & 20.95 & 0.05 \\
2010 & Aug & 19.25 &   32.44 & \zps & 21.02 & 0.05 \\
2010 & Aug & 31.24 &   42.69 & \zps & 21.13 & 0.06 \\
2010 & Sep &  3.24 &   45.25 & \zps & 21.32 & 0.07 \\
2010 & Sep &  6.24 &   47.82 & \zps & 21.47 & 0.09 \\
2010 & Sep &  9.23 &   50.38 & \zps & 21.39 & 0.08 \\
2010 & Sep & 12.26 &   52.96 & \zps & 21.25 & 0.10 \\
2010 & Sep & 15.25 &   55.52 & \zps & 21.60 & 0.10 \\
2010 & Sep & 18.23 &   58.07 & \zps & 21.79 & 0.14 \\
2011 & Feb &  3.65 &  176.42 & \zps & 22.66 & 0.23 \\
2011 & Apr & 22.56 &  243.03 & \zps & 22.66 & 0.23 \\
2011 & Apr & 25.55 &  245.59 & \zps & 22.17 & 0.43 \\
2011 & May & 13.46 &  260.90 & \zps & $>$21.58 & \\
2011 & Jun & 12.45 &  286.54 & \zps & 23.58 & 0.66 \\
2011 & Jun & 24.46 &  296.81 & \zps & 22.76 & 0.27 \\
2011 & Jun & 30.32 &  301.82 & \zps & 22.91 & 0.30 \\
2011 & Jul &  3.34 &  304.40 & \zps & 23.06 & 0.34 \\
2011 & Jul &  6.31 &  306.94 & \zps & 22.61 & 0.22 \\
2011 & Jul &  9.31 &  309.51 & \zps & 23.14 & 0.40 \\
2011 & Jul & 12.31 &  312.07 & \zps & 22.70 & 0.24 \\
2011 & Jul & 18.30 &  317.20 & \zps & 23.48 & 0.53 \\
2011 & Jul & 21.29 &  319.74 & \zps & 24.10 & 0.93 \\
2011 & Jul & 24.27 &  322.29 & \zps & 22.63 & 0.30 \\
2011 & Jul & 27.26 &  324.85 & \zps & 22.77 & 0.25 \\
2011 & Aug &  2.26 &  329.98 & \zps & 22.93 & 0.28 \\
2011 & Aug &  5.26 &  332.55 & \zps & 23.08 & 0.39 \\
2011 & Aug & 17.28 &  342.82 & \zps & 23.04 & 0.32 \\
2011 & Aug & 20.31 &  345.42 & \zps & 23.15 & 0.37 \\
2011 & Aug & 23.27 &  347.95 & \zps & 23.34 & 0.42 \\
2011 & Aug & 29.25 &  353.06 & \zps & 23.89 & 0.82 \\
2011 & Sep &  1.24 &  355.62 & \zps & 24.07 & 0.93 \\
\hline
\end{longtable}

\clearpage

\begin{longtable}{llrrcrr}
\caption{\textsl{GALEX} Time Domain Survey Observations \label{galextab}}\\
\hline
\hline
\multicolumn{3}{c}{UT Date} & Phase\footnote{In rest-frame days after the peak on 2010 July 12.31 UT} & Filter & Mag & $\sigma$\\
\hline
2009 & May &  9.52 & -366.62 & $NUV$ & $>$23.72 &  \\
2009 & May & 11.98 & -364.51 & $NUV$ & $>$23.74 &  \\
2009 & May & 13.90 & -362.87 & $NUV$ & $>$23.77 &  \\
2009 & May & 15.82 & -361.23 & $NUV$ & $>$23.80 &  \\
2009 & May & 17.80 & -359.53 & $NUV$ & $>$23.84 &  \\
2009 & Jun & 21.52 & -329.85 & $NUV$ & $>$23.80 &  \\
2009 & Jun & 23.57 & -328.10 & $NUV$ & $>$23.75 &  \\
2009 & Jun & 25.42 & -326.51 & $NUV$ & $>$23.70 &  \\
2009 & Jun & 27.48 & -324.76 & $NUV$ & $>$23.64 &  \\
2009 & Jun & 29.60 & -322.94 & $NUV$ & $>$23.63 &  \\
2009 & Jul &  1.66 & -321.18 & $NUV$ & $>$23.68 &  \\
2010 & May &  3.77 &  -59.46 & $NUV$ & $>$23.77 &  \\
2010 & May &  7.74 &  -56.06 & $NUV$ & $>$23.82 &  \\
2010 & May &  9.86 &  -54.25 & $NUV$ & $>$23.85 &  \\
2010 & Jun & 17.68 &  -21.06 & $NUV$ & 19.47 & 0.07 \\
2010 & Jun & 19.95 &  -19.12 & $NUV$ & 19.41 & 0.08 \\
2010 & Jun & 23.57 &  -16.02 & $NUV$ & 19.18 & 0.04 \\
2011 & Apr & 21.84 &  242.41 & $NUV$ & 21.99 & 0.13 \\
2011 & Apr & 23.89 &  244.17 & $NUV$ & 21.79 & 0.12 \\
2011 & Apr & 25.81 &  245.81 & $NUV$ & 21.70 & 0.12 \\
2011 & May &  8.40 &  256.57 & $NUV$ & 21.88 & 0.11 \\
2011 & May & 10.39 &  258.27 & $NUV$ & 21.85 & 0.12 \\
2011 & May & 12.37 &  259.97 & $NUV$ & 22.01 & 0.16 \\
2011 & Jun &  6.56 &  281.51 & $NUV$ & 22.07 & 0.16 \\
2011 & Jun & 10.68 &  285.02 & $NUV$ & 22.48 & 0.16 \\
\hline
\end{longtable}

\clearpage

\begin{landscape}
\begin{longtable}{rllccccrr}
\caption{Log of Spectroscopic Observations \label{spectab}}\\
\hline
\hline
Phase\footnote{In rest-frame days after the peak on 2010 July 12.31 UT} &
UT Midpoint &
Instrument &
Exp. Time &
Wavelength Range &
Resolution &
Airmass &
Slit P.A. &
Parall. Angle \\
 &  &  & (s) & (\AA) & (\AA) &  & (deg) & (deg)
\\
\hline
$-22$  & 2010-06-16.33 & MMT/Blue Channel & $1800$ & $3433-8655$ & $5.5$ & $1.16$
& $131.6$ & $130.9$ \\
$227$ & 2011-04-03.48 & MMT/Hectospec & $3060$ & $3700-9150$ & $5.0$ & $1.08$ &
Fiber & \empty \\
$254$ & 2011-05-05.47 & MMT/Blue Channel & $1200$ & $3396-8616$ & $5.5$ & $1.21$
& $117.7$ & $117.6$ \\
$255$ & 2011-05-06.42 & MMT/Blue Channel & $1800$ & $3396-8616$ & $5.5$ & $1.12$
& $141.7$ & $141.5$ \\
$358$ & 2011-09-04.23 & MMT/Blue Channel & $1500$ & $3394-8622$ & $5.5$ & $1.78$
& $83.8$ & $83.8$ \\
\hline
\end{longtable}
\end{landscape}


\end{document}